\documentclass[12pt]{article}
\usepackage[english]{babel}
\usepackage[letterpaper,top=2.6cm,bottom=2.6cm,left=3cm,right=3cm,marginparwidth=1.75cm]{geometry}
\usepackage{fancyhdr}
\usepackage{caption}
\usepackage{amsmath}
\usepackage{amssymb}
\usepackage{graphicx}
\usepackage{helvet} 
\usepackage[colorlinks=true, allcolors=blue]{hyperref}
\usepackage{listings}
\usepackage{xcolor}
\usepackage{tabularx} 
\usepackage{longtable}
\usepackage{booktabs} 
\usepackage[skip=3pt plus1pt, indent=20pt]{parskip}
\usepackage{placeins} 
\newcommand{\codefont}[1]{{\sffamily\small #1}}

\title{``Minus-One" Data Prediction Generates Synthetic
Census Data with Good Crosstabulation Fidelity}
\author{William H. Press\\
Oden Institute of Computational Engineering and Science\\
The University of Texas at Austin}

\begin{document}
\maketitle

\begin{abstract}
We propose to capture relevant statistical associations in a dataset of categorical survey responses by a method, here termed MODP, that ``learns" a probabilistic prediction function ${\cal L}$. Specifically, ${\cal L}$ predicts each question's response based on the same respondent's answers to all the other questions. Draws from the resulting probability distribution become synthetic responses. Applying this methodology to the PUMS subset of Census ACS data, and with a learned ${\cal L}$ akin to multiple parallel logistic regression, we generate synthetic responses whose crosstabulations (two-point conditionals) are found to have a median accuracy of $\sim5$\% across all crosstabulation cells, with cell counts ranging over four orders of magnitude. We investigate and attempt to quantify the degree to which the privacy of the original data is protected.
\end{abstract}

\section{Introduction}

This paper proposes a method, possibly new, for capturing the possibly complex structure of causal associations in a large sample of categorical survey data, and then tests the method on the specific problem of generating synthetic census data records. Since the method by itself comes with no mathematical guarantees, it is likely useful only for particular data sets, hence the plan here of both describing the method and demonstrating some degree of its success in the particular application that we now introduce.

\subsection{Microsampling the American Community Survey}
In decennial censuses prior to 2010, about one household in six
received a so-called long form questionnaire with additional
questions, collectively yielding more detailed socioeconomic data
about the population. As a replacement for the long form,
starting in 2005, Census created the American Community Survey (ACS) \cite{ACS}, a
questionnaire sent to a small percentage of households on a
rotating basis and not associated with any particular decennial.
The 2010 and 2020 decennials comprised only short form questionnaires.

As of 2022, the ACS consisted of approximately 100 questions about
each household (``housing records"), and about 70 questions about each
individual person in the household (``person records"). The person
records are augmented by about 50 derived ``recode" questions. Most
questions have corresponding ``allocation flags" that indicate
whether the response was given by the respondent or was imputed by
other means. A large majority of the questions are categorical
(i.e., one response to be chosen from a given list of responses).
Some of the remaining questions, for example regarding income, allow
numerical responses.

ACS data is published as tables aggregating responses in various
geographical subdivisions that have more than a specified minimum
number of responses (varying, but of order one hundred thousand).  Also
released, mostly with single-state resolution, is a Public Use Microdata
Sample (PUMS). \cite{PUMS} The PUMS records are a statistical sample of
individual household or person records, numbering about one percent
of each state's population. Although the PUMS records correspond
to actual responses, various disclosure protection methods are
applied to defeat in principle the reidentification of
individual respondents. Disclosure protection methods may include
any or all of: data swapping, topcoding, aggregation of some responses
(e.g., occupation codes) into broader categories, or noise infusion.

However, as methods of reidentification improve based on algorithmic
developments, the wide availability of large-scale computational
resources, and the commercial availability of related mass personal
data, Census is rightly concerned that the reidentification of PUMS
data may be possible. Or, put differently, Census may be challenged to justify its present disclosure protection methods
as adequate.

Record synthesis, the creation of fictitious records that
somehow embody the statistical properties of the true data without
revealing that data, seem to offer a way out of the disclosure dilemma. Synthetic records would, ideally, connect to no actual true persons and have no privacy concerns. They could be produced and distributed in unlimited quantity.

Are such synthetic records feasible? Clearly, synthetic records cannot have ``all" the statistical properties of the true records: One can posit contrived statistics that are easily inverted to yield exactly the original data, for example a sufficiently large number of linear combinations of the data. At the other extreme, it is trivial to create synthetic data that duplicates the univariate (i.e., unconditioned) answers to all questions---simply draw responses with the univariate probabilities seen in the population. That protects privacy perfectly, but destroys all statistical associations among answers.

A considerable body of work, beginning with Ullman and Vadhan (2010) \cite{Ullman}, proves that, in general, ``perfect" synthetic records, ones that both protect privacy and reproduce all two-point conditionals even approximately, are impossible. The caveat ``in general," however, means the worst case over all possible data sets. Census data in general, and ACS data in particular, can have statistical regularities that allow useful synthetic data to be generated. Indeed, we next propose a particular method for generating synthetic data and hypothesize that the census data do in fact have the statistical property necessary for that method to usefully work. The remainder of this paper tests that hypothesis by direct computer experiment, with surprising success.

\subsection{Minus-One Data Prediction}
\label{sec:modp}

Imagine a survey with $Q$ categorical questions.
Each question's responses can be represented as a one-hot binary row vector, meaning that it has value $1$ in the column corresponding to the chosen response, $0$ in all other columns. A complete response to the survey is a row vector that concatenates these one-hot responses, with exactly one $1$ in each question's columns. We are given a large sample of these complete responses as the rows of a dataset.

Everything in this paper is based on a single idea: Suppose that we can learn a function
${\cal L}_J$ that gives a probabilistic prediction of question $J$'s response conditioned on its knowing the same respondent's responses to all the other questions. Then, in each row of the dataset, we can substitute for question $J$'s one-hot vector a vector of predicted probabilities based on that row's other responses. We can do this for each question separately, producing a complete set of probabilities. For each row and question $J$, ${\cal L}_J$ is a function of all that row's responses {\it except for} question $J$. The same function is applied to each row's different data.

Finally, we can sample from this collection of probabilities, independently for each question and for each row, to produce a new ``synthetic" sample of responses. (In fact, we can do this as often as we like, mapping each original row into its own ``cloud" of synthetic responses.)

We will refer to this process as ``Minus-One Data Prediction" or MODP, because its defining property is that, in each row, each question is predicted without self-prediction, that is, minus itself.

Why does, or why might, this work? Consider the data's two-point conditional, or crosstabulation, probabilities. If (in somewhat loose notation) $a,b,c,d,\ldots$ are the Boolean values in $\{0,1\}$ of different columns of a single respondent, then one cell in the crosstabulation has probability $P(ab)$, the probability of both $a$ and $b$ being true. We can write
\begin{align}
\label{eq:modp}
P(ab) &= P(a|b)P(b)\\
    &= \sum_{cde\ldots} P(a|bcde\ldots) P(b|cde\ldots)P(cde\ldots)\notag\\
    &= \sum_\text{sample} P(a|bcde\ldots) P(b|cde\ldots)\notag\\
    &\approx \sum_\text{sample} P(a|bcde\ldots) P(b|acde\ldots)\notag
\end{align}
Here we use the fact that the sample is, by definition, weighted by $P(cde\ldots)$ (or, for that matter, any other joint probability).
The last line is exactly MODP, since it sets $a$ and $b$ independently in each row (each conditioned on all columns except itself) and then sums over the sample.
The condition for the approximate equality to hold is,
\begin{equation}
    P(b|cde\ldots) \approx P(b|acde\ldots)
\label{eq:condind}
\end{equation}
which just says that $a$ and $b$ are conditionally independent, given $c,d,e,\ldots$. To expect such conditional independence in a dataset might seem counterintuitive. But no: To the extent that many causal entanglements among the survey questions makes each row (loosely speaking) ``overdetermined", then $c,d,e,\ldots$ alone may give nearly as good a prediction of $b$'s probability as if the value of $a$ were also known.

A calculation similar to equation \eqref{eq:modp}, but for $P(abc)$, gives hope for reproducing not just the two-point conditions, but also three-point and higher---each step requiring greater degrees of ``overdeterminedness" in the underlying data. 

An important point is that these predictions are of probabilities, not of a particular realization. So a prediction can be accurate even if an answer is ``unpredictable" based on all the other columns. In fact, this is the {\it most} favorable case, implying full, not just conditional, independence.

The questions to be considered are these: (i) Are the ACS census data's conditional independences good enough to make MODP (equation \ref{eq:condind}) useful? (ii) Even if they are so in principle, can we in practice learn a sufficiently accurate prediction function? And, (iii) If the original sample is private, how well is that privacy protected in the synthetic sample?

It is beyond the scope of this paper to review the many existing approaches to data synthesis, or even the more restricted case of synthesizing categorical data.  A recent overall review with many references is \cite{UN2022}. Bayesian finite population inference has been much studied (e.g., \cite{Mathur2024}). Among various machine-learning approaches, GANs are popular (e.g., \cite{jordon2018}), and
the concepts of Bayesian generative models (e.g. \cite{zhang2014}) and learned autoencoders (e.g., \cite{abay2018}) are well known. 
The basic idea of MODP, which can be viewed as a specific kind of autoencoder, is so simple that it seems likely to have been already invented; but at time of writing we are not aware of any specific reference in the literature.

\section{Simplified Testbed-Use Micro Sample (STUMS)}
\label{sec:STUMS}

For developing a demonstration of principle, short of a full-scale implementation, it is useful to have a testbed data set that can stand in as a surrogate for actual census data. To avoid privacy issues, we base our testbed on PUMS public records.
As a reasonable sample size for experiments on a moderate scale, we take as input
all 2022 PUMS records for the state of Texas, numbering 292,919. PUMS records are weighted, the weights intended to produce a more representative sample of the population, but, for our purposes, we take the unweighted records as testbed ground truth, as if they were sampled from some non-exactly-Texas population. We similarly ignore all allocation flags and take all responses to have equal validity.

As a simplified subset of the PUMS columns, we select 38 typical questions from the personal records, including ones that seem most defining (sex, age, marital and employment status, etc.). Most questions are already categorical. Because the simulation or synthesis of categorical data presents the most challenges, we convert numerical questions among the 38 into categorical ones by sorting the full set of responses into deciles, yielding in each case 10 categorical answers (plus in some cases one or two categories for missing or non-numerical data). We also simplify some questions that have large numbers of categorical answers (e.g., ``weeks worked in previous year") to a smaller number of categories, typically by merging groups of categories by ranges.

We refer to this simplified resulting data set as the Simplified Testbed-Use Micro Sample (STUMS). As shown in the following table, the 38 STUMS questions have in total 233 categorical responses. The STUMS data set is thus a binary matrix of shape $(292919, 233)$
with the property of being one-hot (having a single 1 value per row) in 38 groups of adjacent columns, defined by a length-38 list of the number of columns assigned to each question.

\begin{longtable}{@{} l >{\raggedright\arraybackslash}p{6cm} ccc @{}}
\caption{STUMS Subset of PUMS Questions} \label{Table1} \\
\toprule
Question & Description & \# Cats & Merged & Quantiled \\
\midrule
\endfirsthead
\multicolumn{5}{c}{{Table \thetable\ (continued)}} \\
\toprule
Question & Description & \# Cats & Merged & Quantiled \\
\midrule
\endhead
\midrule
\endfoot
\bottomrule
\endlastfoot
SEX & Sex & 2 & & \\
AGEP & Age & 11 & Yes & \\
RAC1P & Recoded detailed race code & 9 & & \\
NATIVITY & Nativity & 2 & & \\
WAOB & World area of birth & 8 & & \\
CIT & Citizenship status & 5 & & \\
DECADE & Decade of entry & 9 & & \\
MIL & Military service & 9 & & \\
ENG & Ability to speak English & 9 & & \\
HICOV & Health insurance coverage recode & 2 & & \\
SCHL & Educational attainment & 10 & Yes & \\
SCH & School enrollment & 4 & & \\
SCHG & Grade level attending & 6 & Yes & \\
MAR & Marital status & 5 & & \\
MSP & Married, spouse present/spouse absent & 7 &  & \\
MARHT & Number of times married & 4 & & \\
MARHYP & Year last married & 9 & Yes & \\
MARHW & Widowed in the past 12 months & 3 & & \\
MARHM & Married in the past 12 months & 3 & & \\
MARHD & Divorced in the past 12 months & 3 & & \\
FER & Gave birth to child within the past 12 months & 3 & & \\
ESR & Employment status recode & 7 & & \\
WRK & Worked last week & 3 & & \\
WKHP & Usual hours worked per week past 12 months & 11 & & Yes\\
WKWN & Weeks worked during past 12 months & 7 & Yes & \\
JWAP & Time of arrival at work, hour and minute & 11 & Yes & \\
JWMNP & Travel time to work & 11 & & Yes \\
JWTRNS & Means of transportation to work & 8 & Yes & \\
LANX & Language other than English spoken at home & 3 & & \\
LANP & Language spoken at home & 4 & Yes & \\
DIS & Disability recode & 2 & & \\
DEYE & Vision difficulty & 2 & & \\
DEAR & Hearing difficulty & 2 & & \\
DREM & Cognitive difficulty & 3 & & \\
PINCP & Total person's income & 10 & & Yes\\
PERNP & Total person's earnings & 10 & & Yes\\
SSP & Social Security income past 12 months & 12 & & Yes\\
PAP & Public assistance income past 12 months & 12 & & Yes\\
\midrule
Questions: 38 &   \multicolumn{2}{r}{Categories: 233
$\;$}&  & \\
\end{longtable}

In the Supplementary Materials, we define an independent subset of the PUMS data, based on its housing records instead of the person records. Here termed STUMS-H, that subset has 133,016 rows, with 46 questions that expand to 315 binary columns. The Supplementary Materials repeats the full analysis of this paper on that independent data set as an end-to-end check of the methods here developed and tested on STUMS.

\subsection{What Would Ideal Look Like?}

Before attempting to create synthetic data, is is useful to know what a perfect synthetic data set generated from STUMS would look like. Should it have {\it exactly} the same crosstabulation counts as STUMS itself?

No. PUMS, STUMS, and therefore also any synthetic substitute, represents a sample of an underlying much larger population. Any two samples of that population will naturally differ. We can estimate this effect by bootstrap resampling \cite{Efron81} of (here) the STUMS data.

\begin{figure}
\centering
\hspace*{-40pt}
\includegraphics[width=400pt]{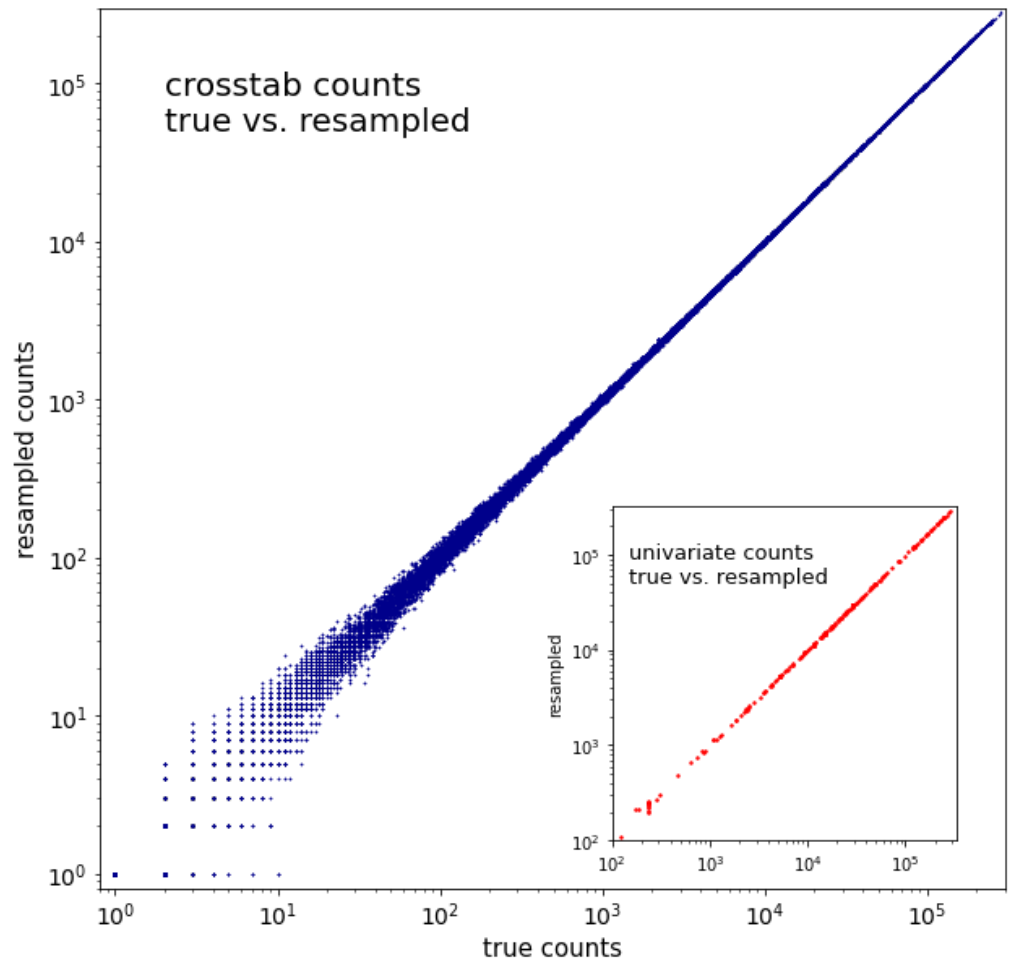}
\caption{\label{fig:perfection}Changes in the univariate and crosstabular counts of the STUMS data set under bootstrap resampling \cite{Efron81} of its 292,919 rows. In the main panel, the discrete lines of dots correspond to $0,1,2,\ldots$ counts in the true or synthetic data. (Zeros are plotted as 1 for visibility.) Achieving a similar scatter with respect to true data is a (likely unobtainable) goal of any method for producing synthetic data---the best we can hope for.}
\end{figure}

Figure \ref{fig:perfection} shows the results of such resampling. It represents the goal, likely unobtainable, of any attempt at creating synthetic data. When, below, we show results actually obtained, it will be useful and sometimes sobering to refer back to this figure.

\section{Applying MODP to STUMS}

The intuition behind MODP (\S\ref{sec:modp}) is that its probabilistic predictions
for questions that don't much depend on other answers will be, in effect,
independent draws from those questions' unconditional distribution; while
predictions for questions with statistical associations to other questions
will be drawn from distributions reflecting those associations.

Privacy protection in this scheme derives from the stochasticity of the draws.
The realized degree of protection is a question discussed below. But
it is worth noting here that, as each true record is transformed to a synthetic one, we can easily quantify, for the probability values actually drawn from, the amount of entropy introduced---say in bits. Information theory tells us that, if $B$ bits of entropy are introduced, then a synthetic record is just one of about $2^B$ roughly equiprobable such records.\cite{shannon} Privacy will depend on how this cloud of possibilities is distributed with respect to the protected, true data. We will measure this.

This general scheme can be implemented with any number of possible
prediction algorithms, each based in some way on a large number of true responses as training data. One might expect the fidelity of statistical
properties of the synthetic data to depend on the statistical subtlety of the
prediction function. Our experiments will be with a particular class of prediction functions.

There is no theoretical guarantee that this method will work at all. That is a matter for experiment, as we now describe.

\subsection{Notation}

We have a set of categorical questions $\{Q\}$
indexed by an uppercase letter, $Q_I \in\{Q\}$
Each question's responses are encoded as a one-hot vector
\begin{equation}
  R_I \in \{0,1\}^{{\cal N}_I}, \qquad \sum_{k=1}^{{\cal N}_I} R_{Ik} = 1  
\end{equation}
where ${\cal N}_I$ denotes the
number of categorical responses to question $Q_I$.
A complete set of responses to all questions, a ``response vector",
is the concatenation of responses to the individual questions,
\begin{equation}
R = \bigl( \bigotimes_{Q_I \in \{Q\}} R_I \bigr) \in \{0,1\}^{\cal N},
\qquad {\cal N} = \sum_I {\cal N}_I     
\end{equation}

We want a learned function  ${\cal L}$ that maps any true response vector $R_\alpha$ to
a set of synthetic responses, denoted $\{S_\alpha\}$,
that, in some manner to be defined, preserves statistical
properties of the $R_\alpha$'s yet protects them from reidentification,
\begin{equation}
{\cal L} : R_\alpha \rightarrow \{S_\alpha\}
\end{equation}
Computationally, the set of $R_\alpha$'s is encoded as a binary matrix of
shape $[N,{\cal N}]$, where $N$ is the number of responses (rows), ${\cal N}$ the total number of categories (columns),
and $\{S_\beta\}$ is a probability distribution over binary matrices of the same shape,
any member of which has rows that are valid synthetic replacements for the corresponding row's $R_\alpha$. We write $k \in Q_J$
as meaning that column $k$ is one of $Q_J$'s responses.

\subsection{Non-Self Predicting Logistic Regression}
\label{sec:NSPLR}

We will try to learn a function ${\cal L}$ made up of one or more ``minus-one logistic maps", each based on an affine map $(L,c)$ from a binary vector to a real vector,
\begin{equation}
(L,c)  : \{0,1\}^{\cal N} \rightarrow (0,1)^{\cal N}    
\end{equation}
and generating probabilities by
\begin{equation}
\qquad P_i = \text{Sigmoid}\left[\sum_{j=0}^{{\cal N} - 1} L_{ij} R_j + c_i\right]    
\end{equation}
Here, $L$ is a square matrix of reals with the special property (``minus-one" or ``non-self-predicting"),
\begin{equation}
 \forall Q_I \in \{Q\} : L_{ij} = 0 \text{ if } (i \in Q_I \text{ and } j \in Q_I)   
\end{equation}
In other words, $L$ has block-diagonal zero ``holes" that prevent any column
$i \in Q_I$ from participating in that question's map.
The vector $c_i$ are arbitrary biases (offsets).
The sigmoid function is taken to be threaded over columns,
\begin{equation}
\text{Sigmoid} : \mathbb{R}^{\cal N} \rightarrow (0,1)^{\cal N} \qquad
\text{Sigmoid}(x_i) = 1/[1 + \exp(-x_i)]    
\end{equation}
Note that $P$ is a vector of (possibly non-normalized) probabilities in $(0,1)$. In general terms, this scheme is a kind of logistic regression, avoiding self-prediction.
\vspace{6pt}

In PyTorch \cite{PyTorch}, the non-self-predicting map can be implemented as
\vspace{3pt}
\begin{lstlisting}[language=Python]
class NonSelfPredictingLayer(nn.Module):
    def __init__(self, in_features, block_starts, out_features):
        super(NonSelfPredictingLayer, self).__init__()
        self.in_features = in_features
        self.out_features = out_features
        self.block_starts = block_starts
        # weight and bias define a linear map:
        self.weight = nn.Parameter(torch.Tensor(self.in_features, self.out_features))
        self.bias = nn.Parameter(torch.Tensor(self.out_features))
        self.reset_parameters()
        self.weight.register_hook(self.gradient_hook)
    def reset_parameters(self):
        # implement a standard initialization protocol
        nn.init.xavier_uniform_(self.weight)
        bound = 1 / torch.sqrt(torch.tensor(self.out_features, dtype=torch.float))
        nn.init.uniform_(self.bias, -bound, bound)
        self.zero_blocks()
    def zero_blocks(self):
        # set to zero weights that would be self-predicting
        with torch.no_grad():
            for start, end in zip(self.block_starts[:-1], self.block_starts[1:]):
                self.weight[start:end, start:end].fill_(0)
    def gradient_hook(self, grad):
        # set to zero gradients of the zero self-predicting weights
        with torch.no_grad():
            for start, end in zip(self.block_starts[:-1], self.block_starts[1:]):
                grad[start:end, start:end] = 0
        return grad
    def forward(self, input):
        # linear map followed by sigmoid mapping to (0,1)
        self.zero_blocks()
        output = torch.matmul(input, self.weight) + self.bias
        return torch.sigmoid(output)
\end{lstlisting}
Here the trick is to both set the non-self-predict diagonal blocks to zero with \codefont{zero\_blocks()} and also to intercept gradient calculations with a \codefont{gradient.hook()} and set those gradients to zero, so that they are not updated.

The simplest use of \codefont{NonSelfPredictingLayer} would be a model like this,
\vspace{3pt}
\begin{lstlisting}[language=Python]
class OneLayer(nn.Module):
    def __init__(self, in_features, block_starts, out_features):
        super(OneLayer, self).__init__()
        self.layer = NonSelfPredictingLayer(in_features, block_starts, out_features) 
    def forward(self, x):
        x = self.layer(x)
        return x
\end{lstlisting}
to be trained with the standard mean square loss function \codefont{torch.nn.MSELoss}. Inputs and targets for training are identical batches of true response matrices $R_{\alpha k}$. There is no such thing as overtraining here: For our purposes, the closer the prediction the better, as long as we can show an acceptable degree of privacy protection.

After training, model output \codefont{probdat} is sampled for conversion from probabilities to categorical synthetic responses,
\begin{lstlisting}[language=Python]
from torch.distributions.categorical import Categorical # convenient built-in!
def instantiate(probdat, block_starts) :
    log2 = np.log(2.)
    instant = torch.zeros_like(probdat)
    entropies = torch.zeros(probdat.shape[0],device=device) # informational
    for k in range(len(block_starts)-1) : # for each question
        qprobs = probdat[:,block_starts[k]:block_starts[k+1]])
        qprobs /= torch.sum(qprobs,axis=1,keepdim=True) # normalize probabilities
        entropies -= torch.sum((qprobs + 1.e-10)*torch.log(qprobs + 1.e-10),axis=1) / log2
        dist = Categorical(qprobs)
        samples = dist.sample() # sample from qprobs
        # set binary output:
        instant[torch.arange(instant.shape[0]), samples + block_starts[k]] = 1
    return instant, entropies
\end{lstlisting}

\begin{figure}
\centering
\hspace*{-40pt}
\includegraphics[width=400pt]{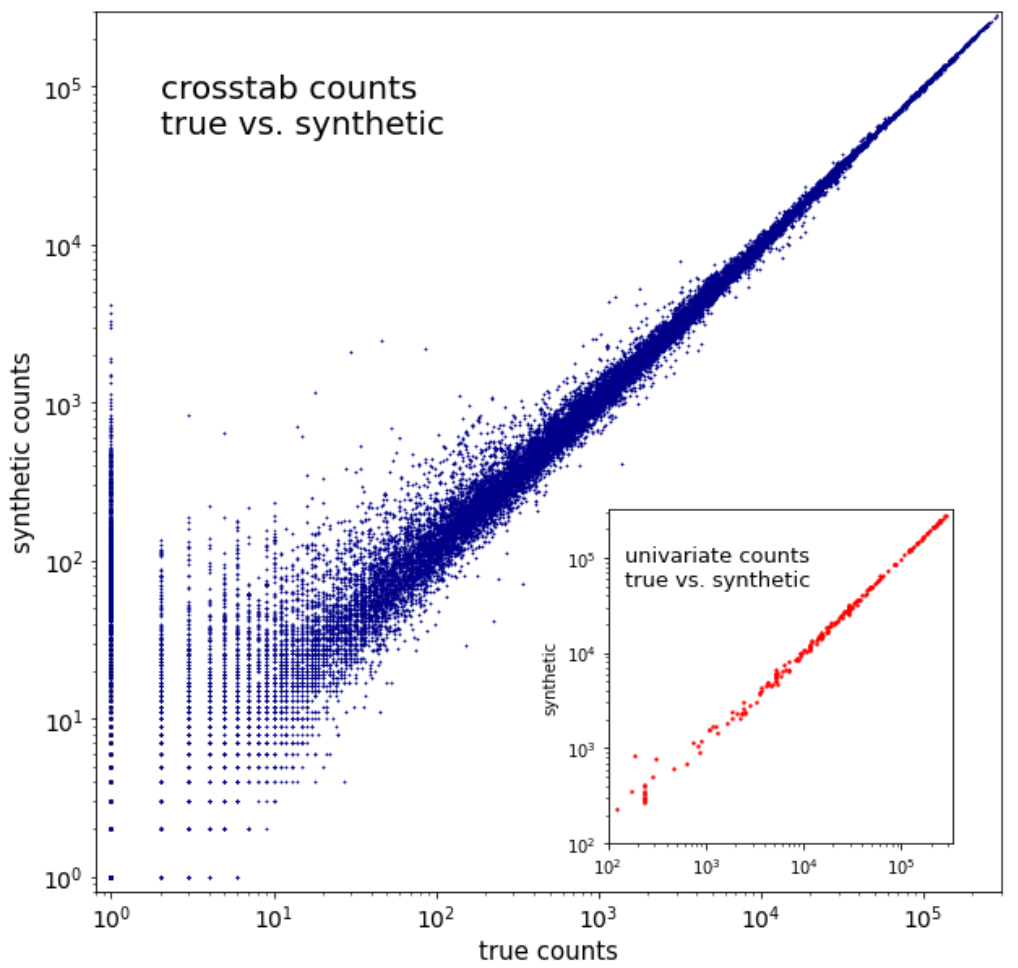}
\caption{\label{fig:oneblade}Univariate and crosstabular predictions of a primitive trained model with a single \codefont{NonSelfPredictingLayer} and loss function \codefont{torch.nn.MSELoss}. Without any direct knowledge of the data's crosstabulations, the model nevertheless reproduces crosstabulation counts to a considerable degree. Visible are this model's deficiencies in populating true structural zeros and cells with small numbers of true counts. More complex models, below, will do much better.}
\end{figure}

Notice that the above code has no direct knowledge of the univariate or crosstabular probabilities of the data. It is only trying to predict probabilities of individual responses in individual response rows, minimizing its probabilities' mean square deviation from the realized values of 0 or 1. That said, the model does surprisingly well at approximating most univariate and crosstabular counts right out of the box, as is seen in Figure \ref{fig:oneblade}. The model is poor on crosstab counts less than $\sim 100$ and especially poor on so-called structural zeros, crosstabulation cells with always zero counts because they represent impossible combinations of answers. Instead of zero probability, the model assigns counts ranging from $1$ to $~10^3$ (out of $\sim 3\times 10^5$), seen in the vertical line of dots at the left of Figure \ref{fig:oneblade}.

We now proceed to more complicated models that do significantly better in all respects.

\section{Multi-Blade Prediction}
\label{sec:multiblade}

Most neural net applications would at this point generalize to add further layers to the model in sequence. That is forbidden here, because it would violate no-self-prediction: Information from question $I$ would pass through the prediction of question $J$ by one layer and then back to $I$ again in the next.

Instead what we can do is to add copies of \codefont{NonSelfPredictingLayer} as parallel ``blades" sharing the same input, and then combine their outputs with trainable weights. Consider this model,
\begin{lstlisting}[language=Python]
class MultiBladeModel(nn.Module):
    def __init__(self, in_features, in_block_starts, out_features, reduced_features, num_blades):
        super(MultiBladeModel, self).__init__()
        self.in_features = in_features
        self.in_block_starts = in_block_starts
        self.out_features = out_features
        self.reduced_features = reduced_features
        self.num_blades = num_blades
        self.blades = nn.ModuleList([NonSelfPredictingLayer(in_features, in_block_starts, out_features) for _ in range(num_blades)])
        self.weight_calculator = WeightCalculator(in_features, reduced_features, num_blades)
        self.sigmoid = nn.Sigmoid()
    def forward(self, x):
        outputs = [blade(x) for blade in self.blades]
        weights = self.weight_calculator(x)
        combined_output = torch.stack(outputs, dim=-1) * weights.unsqueeze(1)
        final_output = combined_output.sum(dim=-1)
        return final_output

class WeightCalculator(nn.Module):
    def __init__(self, in_features, reduced_features, num_outputs):
        super(WeightCalculator, self).__init__()
        self.fc1 = nn.Linear(in_features, reduced_features)
        self.relu = nn.ReLU()
        self.fc2 = nn.Linear(reduced_features, num_outputs)
    def forward(self, x):
        x = self.fc1(x)
        x = self.relu(x)
        x = self.fc2(x)
        weights = torch.softmax(x, dim=1)
        return weights
\end{lstlisting}
This model has \codefont{num\_blades} copies of the basic \codefont{NonSelfPredictingLayer} with separately trainable individual weights. Their output probabilities are combined as a weighted average with \codefont{num\_blades}\ scalar weights. The weights are are a function of each input row and are particular to that input. The function is fixed, but learned in the training phase. In the above example, the weights are calculated as two fully connected linear layers, with a ReLU non-linear activation between them---this, a conventional mini-classifier. The number of intermediate states is a parameter; typically $\sim 10-20$.

If the weights depend in each instance on all the input columns how does this avoid self-prediction? The answer is that the weights are scalars, each one applied across all the columns of its non-self-predicting \codefont{NonSelfPredictingLayer} output. So, training favors the best weights for all columns, no one column in particular.

\begin{figure}
\centering
\includegraphics[width=300pt]{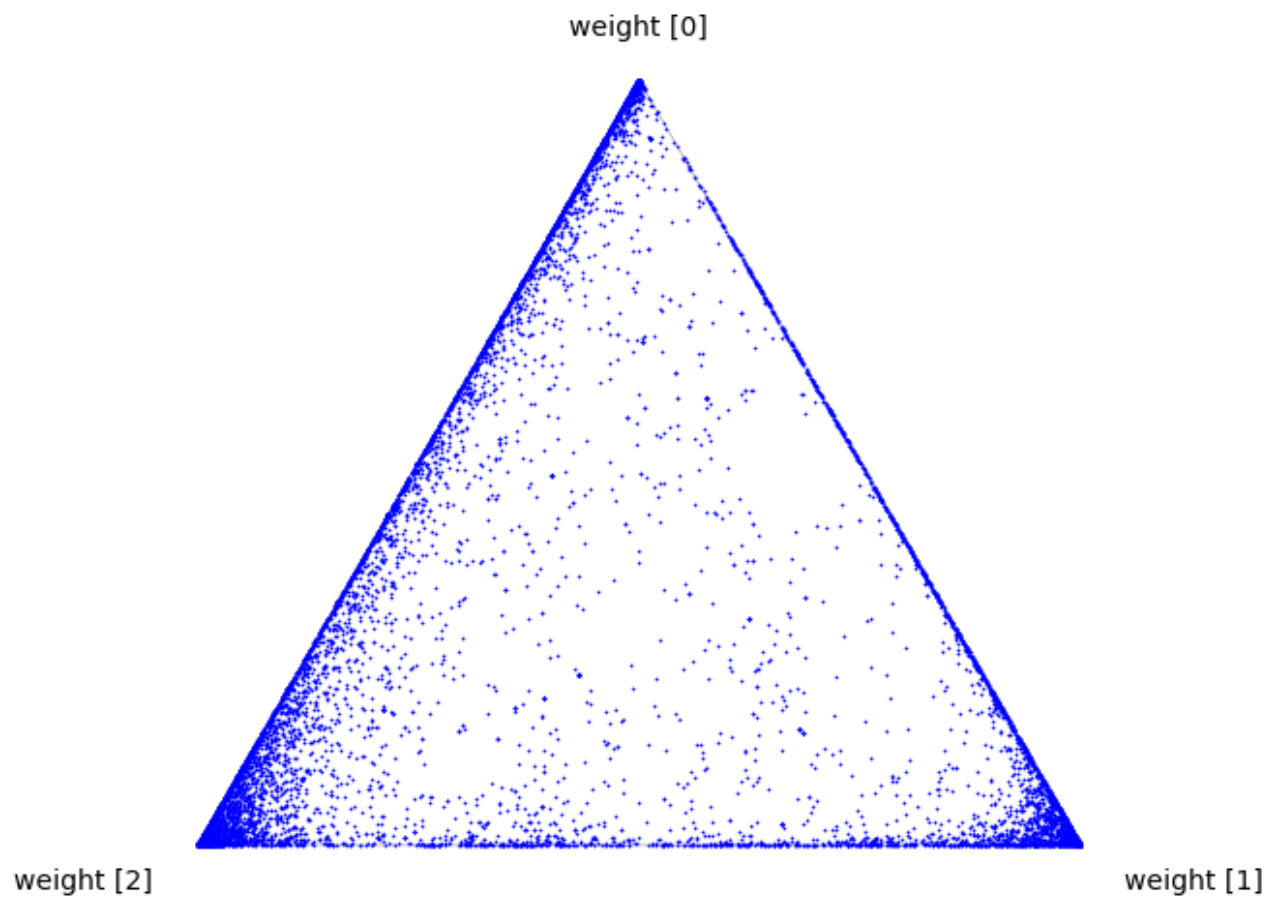}
\caption{\label{fig:triangleplot}For a trained three-blade prediction model, the weights of individual
records are shown as a triangular scatter plot. The concentration of occurrences at vertices and along edges, and the sparsity of points in the interior, suggest that the blades have learned to specialize in predicting different sub-populations of records.}
\end{figure}

The intuition behind this arrangement is a trivial kind of self-attention.\cite{attention} If the input responses contain a few statistically distinct sub-populations, then the model has at least the possibility of optimizing different blades for different sub-populations and training \codefont{WeightCalculator} to recognize in which sub-population an input row lies. Jumping ahead to results more fully explained below, Figure \ref{fig:triangleplot} gives some evidence that this intuition is realized.

The figure plots the weights that a three-blade model uses, after training, in predicting individual rows. One sees that almost all predictions are dominated by either a single blade (vertices of triangle) or a linear combination of two blades (edges of triangle). On the contrary, if the blades were unspecialized, one would expect to see the center of the triangle more heavily populated.

\FloatBarrier
\subsection{Multiblade Training}

We have experimented on models with 1, 3, 5, and 12 blades. Even the largest of these is tiny by modern standards, having only $\lesssim 10^6$ trainable weights. On a single consumer-grade GPU, training times range from minutes for 1 blade to hours with 12. As for all neural nets, training is as much art as science, and one can never be sure that an optimum has been reached.

Our best results are obtained by doing some initial training epochs with the loss function \codefont{torch.nn.MSELoss}. This trains rapidly to roughly reproduce the desired
crosstabulations on a log-log plot (cf. Figure \ref{fig:oneblade}). We then change the
loss function to one whose target is the desired collective crosstabulation rather
than row-by-row prediction of the input, with a figure of merit that is cognizant of expected
count statistics.

We reason as follows:
Suppose we have two binomial draws (a true and a synthetic), $n_{1,2}$ out of $N_{1,2}$ and want to know if
they have a common probability. A standard test \cite{NIST} is the statistic
\begin{equation}
\label{eq:zval}
z = \frac{\hat p_1-\hat p_2}{\sqrt{\hat p (1-\hat p)\left( \frac{1}{N_1} + \frac{1}{N_2}\right)}},
\qquad \hat p_1 = \frac{n_1}{N_1}, \quad \hat p_2 = \frac{n_2}{N_2},\quad \hat p = \frac{n_1+n_2}{N_1+N_2}    
\end{equation}
which is asymptotically normally distributed. This motivates the loss function,
\begin{lstlisting}[language=Python]
def zval_loss_function(output, target, jcbeg):
    cross_output = torch.matmul(output.T,output) + 0.01 # add fractional pseudocount
    cross_target = torch.matmul(target.T,target) + 0.01
    Nt = target.shape[0]
    No = output.shape[0]
    Pt = cross_target / Nt
    Po = cross_output / No
    Pooled = (cross_target + cross_output) / (Nt + No)
    Var = Pooled * (1. - Pooled) * (1/Nt + 1/No)
    zsqvalues = (Pt - Po)**2 / (Var + 1.e-5) # don't let variance get too small
    for start, end in zip(jcbeg[:-1], jcbeg[1:]):
        zsqvalues[start:end, start:end].fill_(0)
    loss = torch.mean(zsqvalues)    
    return loss
\end{lstlisting}
Of course this returns a loss that is not really the sum of independent z-squared values:
First, the columns $k \in K$ within any question $K$ are highly correlated, because they are one-hot. Second, \codefont{target} contains not count data, but probabilities. Hence, third, at the very least we need to mask out the meaningless block-diagonal crosstabulations within individual questions. The code does this. (Not to mention that the analytical formula is only asymptotic!)

One might also ask: Why not train for exactly the loss-function that we really want, namely agreement between the input data and a generated synthetic dataset? The answer is of course that a valid synthetic dataset is integer binary, so implies no gradient on which to descend. And, specifically, the synthetic dataset we do generate involves stochastic choices between 0 and 1, also implying no gradient (in fact, no causal back-calculation of any kind). 

\begin{figure}
\centering
\hspace*{-40pt}
\includegraphics[width=400pt]{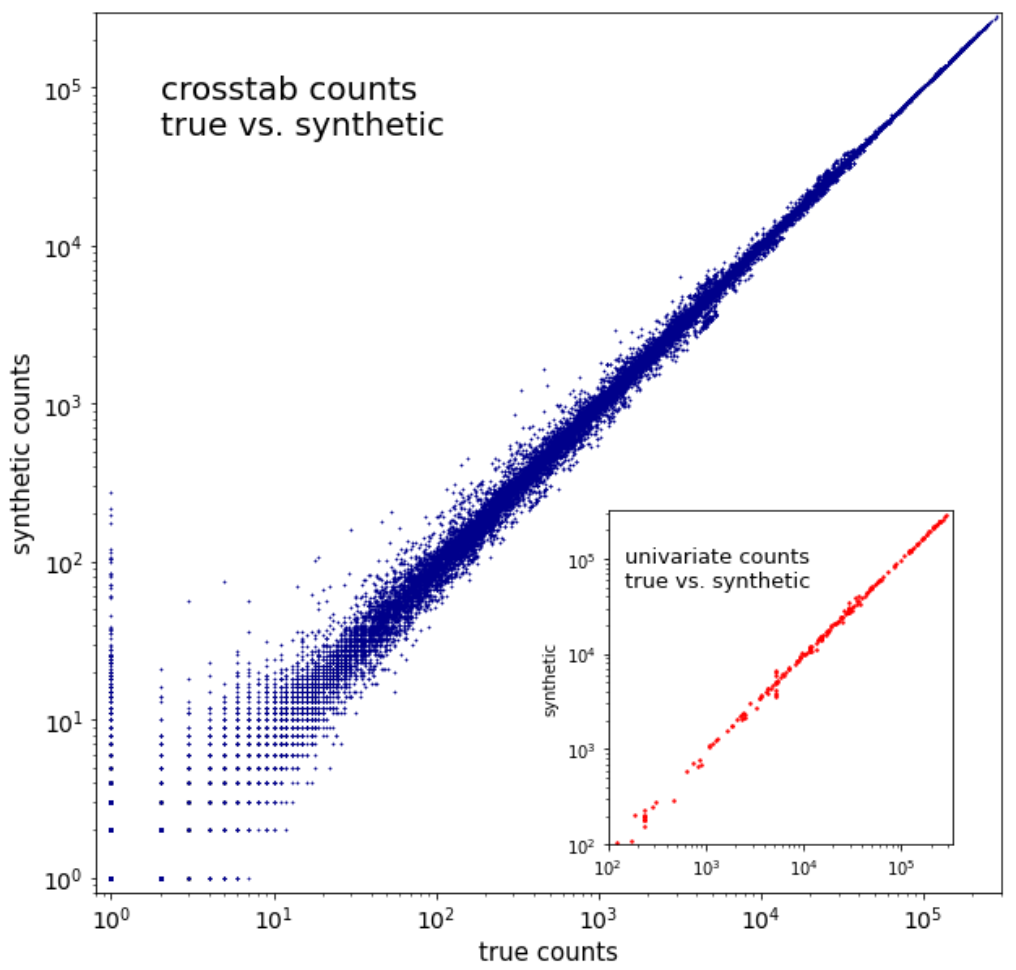}
\caption{\label{fig:fiveblade}Results for a 5-blade model trained with the loss function
\codefont{zval\_loss\_function}. Other than small tweaks (see text) this is our best-performing model. (Compare to hoped-for ideal in Figure \ref{fig:perfection}.)}
\end{figure}

In favor of \codefont{zval\_loss\_function} is that (i) it drives output towards target with a single global minimum, (ii) it has some awareness of count statistics, and (iii) it seems to work well.

Figure \ref{fig:fiveblade} shows results for a 5-blade model trained as described. A small number of crosstabulation cells with too-large synthetic counts continue to catch the eye, and there is still a problem with structural zeros. But, otherwise, the univariate and crosstabulation counts are reproduced with remarkable fidelity.

\FloatBarrier

\subsection{Structural Zeros}
\label{sec:struczeros}

Structural zeros are crosstabulation cells that are zero not just due to sampling, but because they are impossible combinations of responses, for example, a married four-year-old. While few-count violations of a few structural zeros may not be statistically important, their existence can undermine user confidence in the synthetic dataset.

\begin{figure}
\centering
\hspace*{-40pt}
\includegraphics[width=400pt]{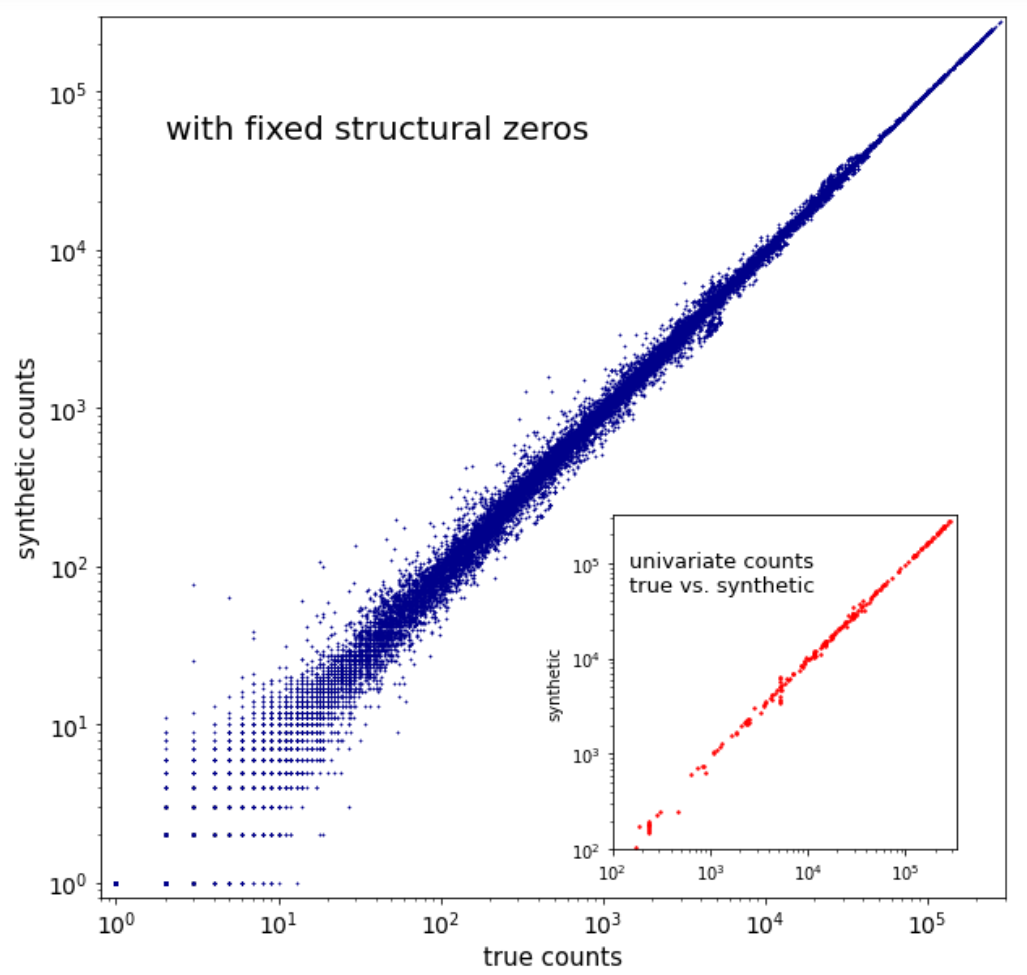}
\caption{\label{fig:tweaks}The model shown in Figure \ref{fig:fiveblade} is here tweaked to remove structural zeros. Performance is otherwise comparable.}
\end{figure}

We have found no elegant solution to the problem of suppressing structural zeros in MODP. But there is an easy brute-force solution: Remove all rows that would generate a count in any structural zero crosstab cell. This works if there are not too many of these. For the output shown in Figure \ref{fig:fiveblade}, it amounts to only $\sim 3$\% of the rows. Figure \ref{fig:tweaks} shows the result.  Quantitative measures of accuracy (Table \ref{tab:LogDevAcc}, below) are essentially unchanged.

\subsection{Randomized Responses}
\label{sec:RR}

The method of randomized responses was developed in the 1960s by Warner \cite{Warner1965}, Greenberg \cite{Greenberg1969}, and others, as a technique for privacy protection in categorical surveys. Categorical responses to questions are recorded truthfully with probability $p$, or as a uniform random choice with probability $1-p$. Any particular respondent has thus a measure of deniability on any question, while the experimenter can calculate true aggregate probabilities after that fact.

\begin{figure}
\centering
\hspace*{-40pt}
\includegraphics[width=260pt]{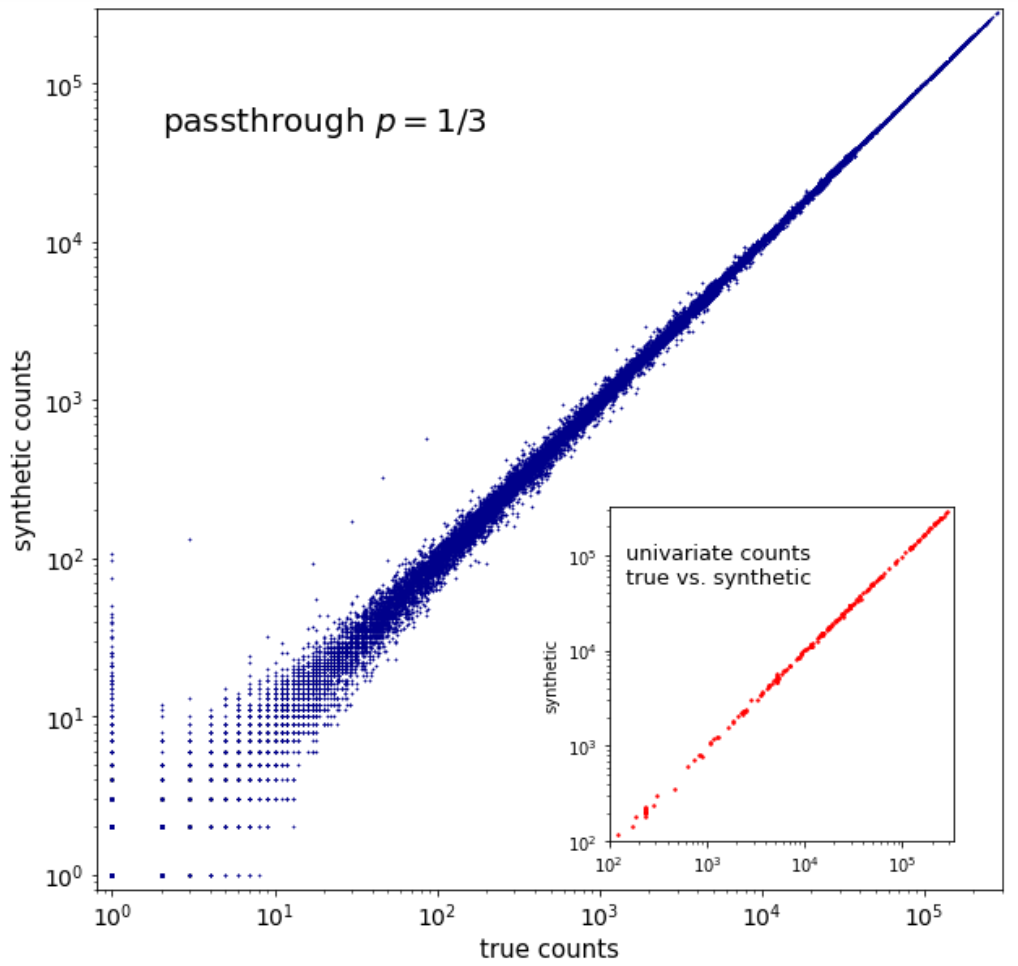}\includegraphics[width=260pt]{randomizedA.png}
\caption{\label{fig:randomized}The accuracy of the model in Figure \ref{fig:fiveblade} is here increased by the method of randomized responses: True responses to each question are passed through to the corresponding synthetic row with probability $p=1/3$ (left) or $p=1/2$ (right) . The increased accuracy is at a cost of some statistical decrease in privacy.}
\end{figure}

In its original application, the method increases privacy but decreases accuracy. Here, it is the opposite: On any response, we can pass through the true value with some fixed probability $p$, versus a draw from the synthetic output's distribution with probability $1-p$. The respondent now {\it loses} some measure of privacy because, question-by-question, there is a greater probability (though never certainty) that the synthetic answer mirrors the actual response. The benefit is that the statistical accuracy of the synthetic data is increased. Figure \ref{fig:randomized} shows the result of doing this with pass-through probabilities $p=1/3$ and $p=1/2$. Structural zeros could be eliminated if desired, as above.

\subsection{Post-Processing}
\label{sec:postp}

A synthetic dataset can be processed after the fact in various ways to improve its accuracy, privacy, or perhaps both. Fixing structural zeros (\ref{sec:struczeros}) is one example. 

Another example is provided by the fact that the probabilistic model can be instantiated multiple times from the same input data, so that multiple synthetic instances are available for each input row. We generate the final output by choosing one of these instances, according to some criterion. This of course introduces a statistical bias; but if the bias is towards a desired end (e.g., better crosstab accuracy), then it is not objectionable.

We have obtained good results by generating just two instances for each row. We define a row-wise loss function by multiplying the (binary valued) crosstab of each row by a weighting function that is large for less accurate crosstab cells, and then summing. When this loss function is larger than a threshold value, we output that row's second instance instead of its first (with no check on the second instance's loss function). This post-processing significantly improves the accuracy of the synthesis, see Table \ref{tab:LogDevAcc} below.

There are other possibilities for post-processing, including different kinds of shuffling steps. One can swap answers between different rows, choosing only swaps that increase crosstab accuracy by a defined criterion. This is slow, but can be significantly parallelized on a GPU. Results on this will be reported elsewhere. 

\begin{table}[h]
\centering
\caption{Fractional (i.e., Log) Accuracy of Various MODP Variants}
\vspace{-6pt}
\label{tab:LogDevAcc}
\begin{tabular}{@{}lcccl@{}} 
\toprule
& \multicolumn{4}{c}{accuracy over $233\times (233+1)/2$ crosstab cells} \\
\cmidrule(l){2-5}
model* & median & mean absolute & r.m.s. & comment\\ 
\midrule
model\_1\_1 & 0.116 & 0.596 & 1.156 & simple one-blade (Fig.~\ref{fig:oneblade})\\
model\_3\_10 & 0.046 & 0.227 & 0.580 & could train better? \\
model\_5\_15 & 0.046 & 0.164 & 0.382 & best model (Fig.~\ref{fig:fiveblade})\\
model\_5\_15 & 0.047 & 0.145 & 0.304 & struct.~zeros fixed (Fig.~\ref{fig:tweaks}) \\
model\_12\_24 & 0.066 & 0.190 & 0.419 & could train better? \\
model\_5\_15 & 0.037 & 0.146 & 0.354 & post-processed (\S \ref{sec:postp})\\
model\_5\_15 (R.R.) & 0.027 & 0.126 & 0.338 & \S\ref{sec:RR} ($p=1/3$)\\
model\_5\_15 (R.R.) & 0.023 & 0.112 & 0.308  & \S\ref{sec:RR} ($p=1/2$)\\
ideal goal & 0.012 & 0.058 & 0.139 & resample of STUMS data (Fig.~\ref{fig:perfection})\\
\midrule
\multicolumn{5}{l}{*``model\_$x$\_$y$" denotes a model with $x$ blades and $y$ reduced features}\\
\bottomrule
\end{tabular}
\end{table}

\FloatBarrier

\section{Accuracy Metrics}

\begin{figure}
\centering
\includegraphics[width=400pt]{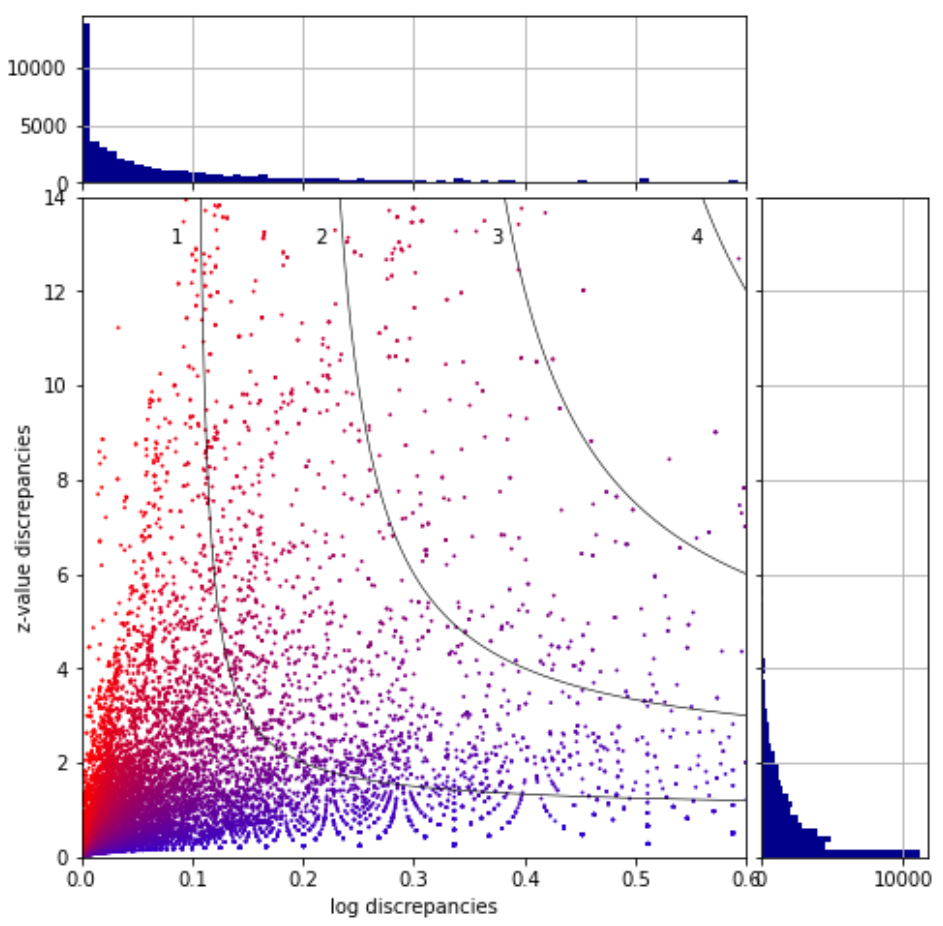}
\caption{\label{fig:zvslog}Accuracy of a trained 5-blade model by two figures of merit. Each dot represents one crosstabulation cell. Its number of counts relative to the true data can be parameterized either by the log discrepancy (roughly, fractional accuracy) or by its statistical z-value discrepancy. For the data shown, the median log discrepancy is 0.05, the median z-value is 0.87. Colors vary from cells with the smallest numbers of counts (blue) to largest numbers (red). The ``suspension bridge" artifacts are from cells with small integer numbers of counts. Black hyperbolic lines are the blended figure of merit, equation \eqref{eq:blended}, see text.}
\end{figure}

Above, in Figures \ref{fig:oneblade}, \ref{fig:fiveblade}, and \ref{fig:tweaks}, we assessed the accuracy of the synthetic data only by eye. Here we can be more quantitative, recognizing that there is more than one possible choice of metric, and that choice will depend on the intended application, which we may not know in advance.

One possible metric is the z-value statistic $z$ (equation \ref{eq:zval}), either individually for each cell or else taking the mean absolute value over all cells in the crosstabulation. Because the cells are not independent, this mean will not have its usual $N(0,1)$ distribution; but it will still give a qualitative answer to the question, ``how statistically significantly does the synthetic data differ from the true," with a value of order unity for good agreement. This metric will not overly punish deviations in cells with very small numbers of counts that are therefore not well determined statistically.

A different possible metric is the absolute logarithmic deviation
\begin{equation}
\label{eq:fracacc}
    d = \left| \log\left(\frac{C^\text{syn}_j+c}{C^\text{true}_j+c}\right)\right|
\end{equation}
where the $C$'s are cell counts and $c$ is a regularizing number of pseudocounts \cite{pseudocount}, typically $0.5$ or $1$. As for $z$, this can be calculated either for each individual cell, or else as an average is over all crosstab cells. This metric will not overly punish cells for having large numbers of counts, where even small fractional discrepancies are highly statistically significant.

A purely heuristic figure of merit is a compromise combination of $z$ and $d$, related to their harmonic mean,
\begin{equation}
\label{eq:blended}
    \text{f.m.} \equiv 2 \left( \frac{d_0}{|d|} + \frac{z_0}{|z|} \right)^{-1}
\end{equation}
where $d_0$ is chosen as ``accurate enough, even if statistically different" and $z_0$ is chosen as ``statistically close enough, even if not accurate". We will generally take $z_0 = 1$ and $d_0 = 0.1$, so that one unit of this f.m.~implies a 10\% fractional accuracy or 1 standard deviation statistical discrepancy from the true data, or a compromise worse on one but better on the other.

\subsection{Distribution of Accuracies of Individual Crosstab Cells}

Figure \ref{fig:zvslog} shows for the individual cells all three of the above metrics for the trained 5-blade model shown above in Figure \ref{fig:fiveblade}. The median log discrepancy $d$ is 0.047, that is, about 5\% fractional accuracy. The median z-value, as defined above, is 0.87. The blended figure of merit (Equation \ref{eq:blended}) is shown as the hyperbolic lines. As expected, cells with the smallest counts (shown as blue) tend to have the largest log discrepancies, even as their statistical significance is acceptable, while cells with the largest numbers of counts (red) are the reverse.

\begin{figure}
\centering
\includegraphics[width=350pt]{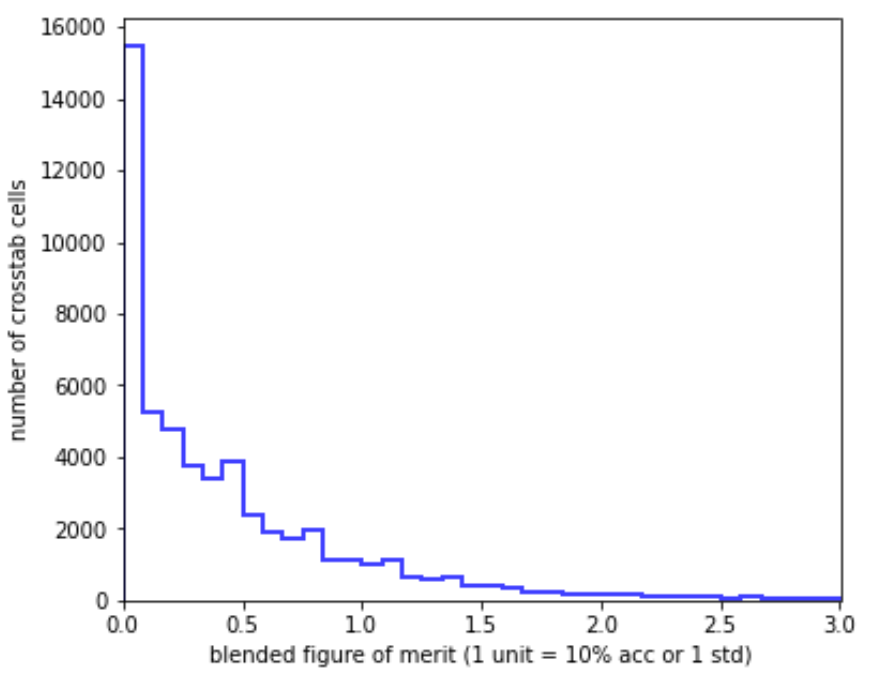}
\caption{\label{fig:fmhistogram}Distribution of blended figures of merit of crosstab cells, same as delineated by the hyperbolic curves in Figure \ref{fig:zvslog}. One abscissa unit signifies 10\% fractional accuracy or 1 standard deviation discrepancy from the true data or a compromise between the two.}
\end{figure}

Figure \ref{fig:fmhistogram} shows the histogram of values of the blended metric $b$, making more visible the large concentration of points close to the origin in Figure \ref{fig:zvslog} and the long-tailed distribution.

Note that the accuracy measures $z$ and $d$ scale differently with sample size. The values shown are for a full synthesis of 292919 synthetic rows (or 288605 if structural zeros are removed). A smaller sample would have smaller (better) $z$ values, and (up to sampling variability) the same $d$ values. 

\begin{figure}
\centering
\includegraphics[width=450pt]{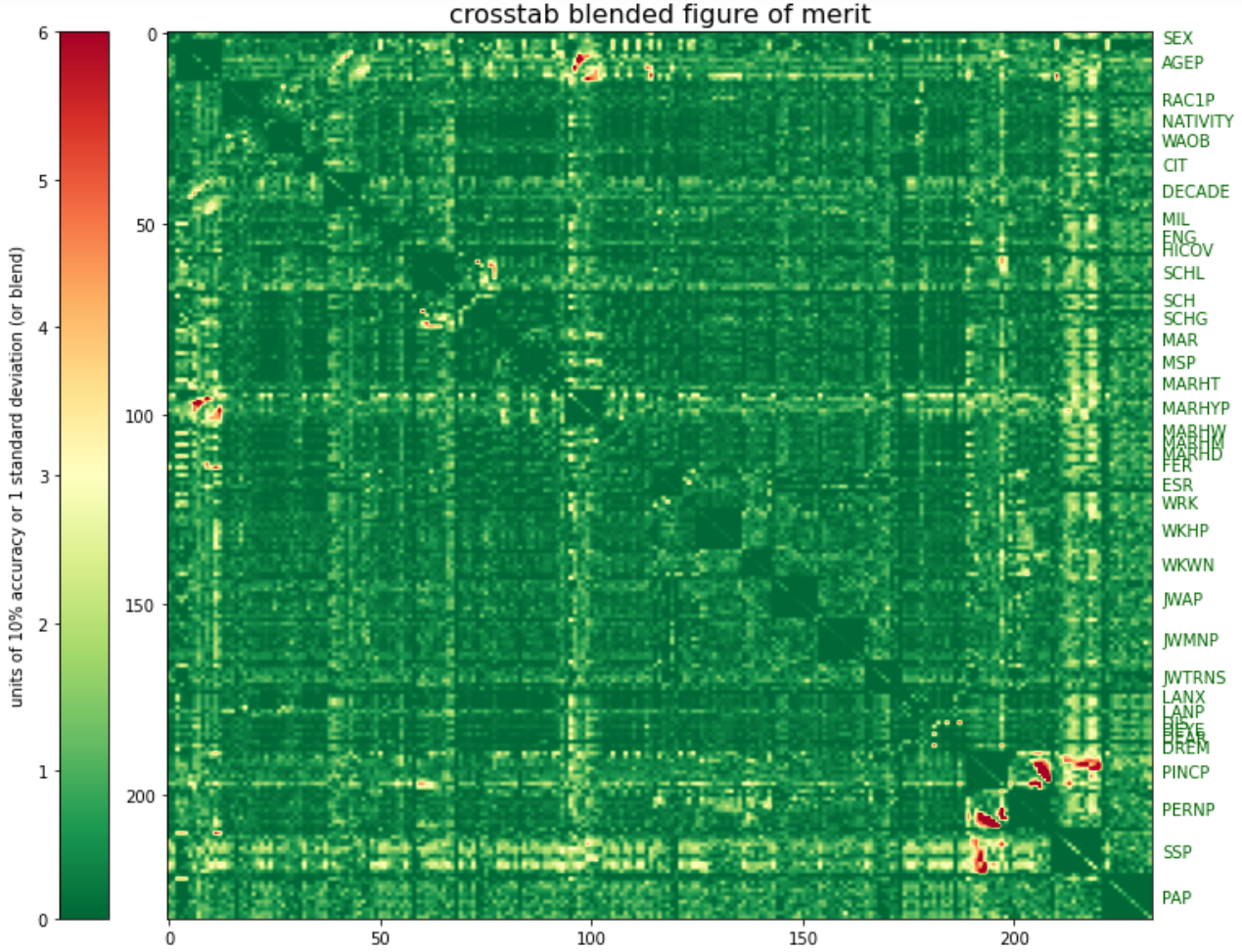}
\caption{\label{fig:plaid}Blended figure of merit for comparing synthetic to real data for every cell in the crosstabulation. On this scale, any shade of green is ``good": a green value 1 indicates 10\% fractional accuracy or 1 standard deviation discrepancy from the true data or a compromise between the two as shown in the hyperbolic lines in Figure \ref{fig:zvslog}. A light-green value 2 indicates 20\% or 2 standard deviations, etc.}
\end{figure}

Figure \ref{fig:plaid} plots the values of the blended figure of merit $b$ as a heatmap of the full $233\times 233$ crosstab. The overall green cast reflects the concentration of points close to the origin in Figures \ref{fig:zvslog} and \ref{fig:fmhistogram}, but one also sees where the model had the most difficulty, namely on the questions about income and age that were most coarsely binned (e.g., by deciles) in our simplified STUMS dataset. Whether the same issues would occur at scale with the full PUMS and, if so, whether they could be mitigated, is beyond our scope.

\subsection{Median, Mean, and R.M.S. Fractional Accuracy across Crosstab Cells}
\label{sec:aggmeas}

Across all the $233\times (233+1)\,/\,2$ cells in the crosstabulation, we can summarize a model by the median, mean-absolute, or root-mean-square (r.m.s.) value of $d$ (Equation \ref{eq:fracacc}). Since the distribution of accuracies is long-tailed, we expect these to be from smallest to largest in the order just mentioned. Table \ref{tab:LogDevAcc}, above, showed all three measures across a variety of MODP models.

One sees in the Table that our best models have median fractional errors of as small as a few percent, albeit with mean absolute errors on the order of 10\% and larger r.m.s.~fractional errors of several tens of percents.

\FloatBarrier
 
\subsection{Varying the Parameters}

Before settling on the trained 5-blade model shown in the figures above and Table \ref{tab:LogDevAcc}, we experimented with 1-blade (Figure \ref{fig:oneblade}), 3-blade, 5-blade, and 12-blade models, and also with variants of the nonlinear \codefont{WeightCalculator} (\S \ref{sec:multiblade}), varying the value of \codefont{reduce\_features}, or adding a third linear layer and second \codefont{ReLU}. Although with 3, 5, or 12 blades we readily obtained results roughly comparable to the 5-blade model shown, none were qualitatively better. We take this as evidence, albeit weak, that we are learning the prediction functions ${\cal L_J}$ (\S \ref{sec:modp}) about as well as is possible, and that the accuracy of MODP is here intrinsic in the data, as by equation \eqref{eq:condind} above. However, this can be only conjectural.
 
\begin{figure}
\centering\hspace*{-40pt}
\includegraphics[width=300pt]{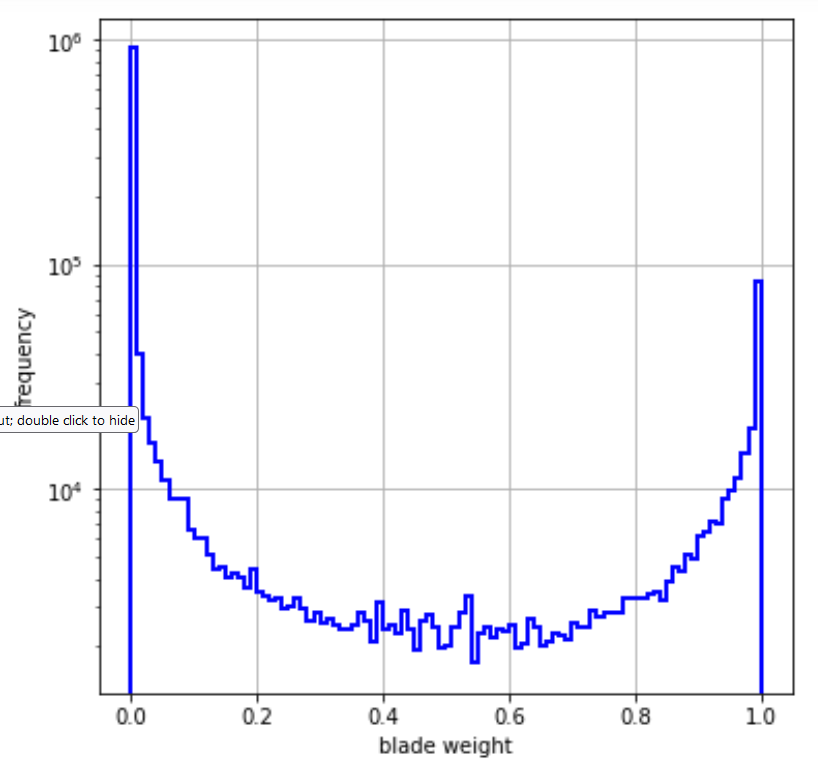}
\caption{\label{fig:fiveweights}The trained 5-blade model assigns row-dependent weights to its blades predominantly with values near to 0 or 1, evidence that it is able to recognize, and optimize for, statistically distinct subpopulations. Note logarithmic $y$-axis.}
\end{figure}

In Figure \ref{fig:triangleplot}, we showed that the 3-blade model actually made use of its learned weight functions in a nontrivial way, assigning strong weights to different blades for different rows. It is not easy to visualize the distribution of weights over a five-dimensional simplex, but we can at least look at the distribution of all weights, shown in Figure \ref{fig:fiveweights}. One sees that weights close to zero or one are strongly favored (note log scale). A similar concentration is seen for the 12-blade models tried.

\begin{figure}
\centering
\includegraphics[width=350pt]{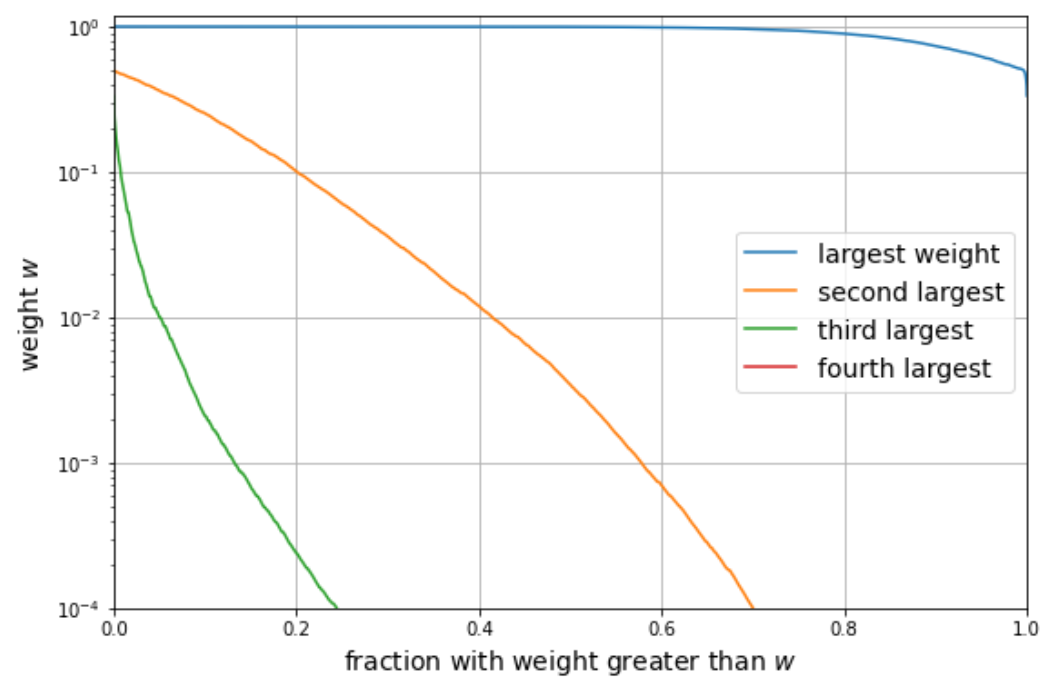}
\caption{\label{fig:weightsdieoff}Cumulative distribution function of the largest weights, second-largest, and so forth, for a trained 5-blade model. As examples of reading this figure, more than 90\% of rows have a largest weight greater than about 0.7, and only about 5\% of rows have a third-largest weight greater than $0.01$.}
\end{figure}

Figure \ref{fig:weightsdieoff} makes the same point in a different way. We show for the trained 5-blade model, the cumulative distribution function of the largest weights, second-largest, and third-largest. (Fourth-largest weights are all effectively zero.)

\begin{figure}
\centering
\includegraphics[width=400pt]{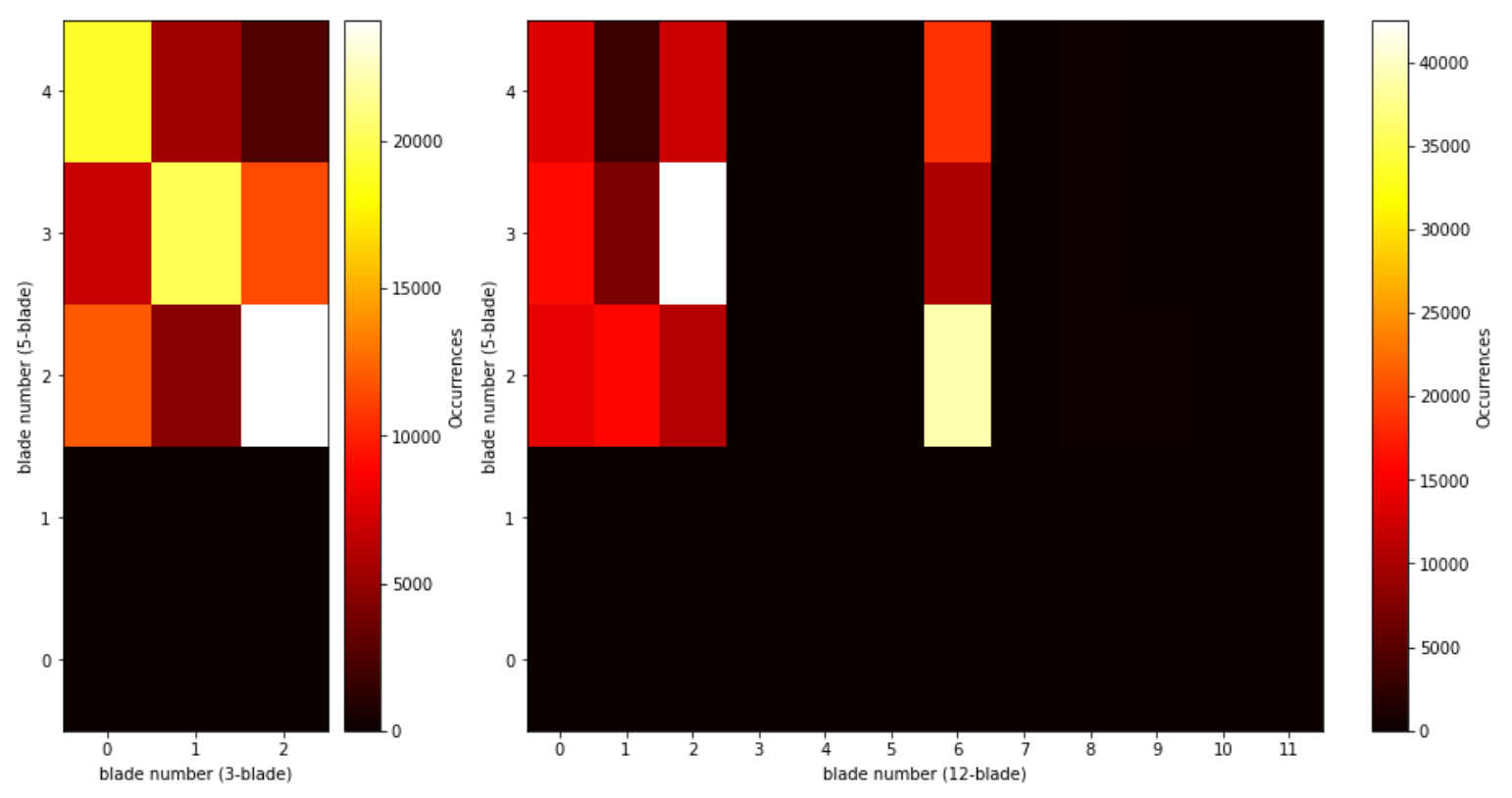}
\caption{\label{fig:crossblade}Heatmap showing the co-assignments after training of rows to blades in 3- and 5-blade models (left), and 5- and 12-blade models (right). The 5-blade model is the same in the left and right panels. Black rows or columns show blades that, after training, are not used at all. The yellow and white entries show that the 5- and 12-blade models tend to recapitulate the same assignment of rows to blades as the 3-blade model.}
\end{figure}

\subsubsection{Blade Specialization}
In fact, as shown in Figure \ref{fig:crossblade}, the 5-blade model, after training, only makes use of three of its blades, and these in individual close correspondence to those of trained 3-blade model. The figure also shows a trained 12-blade model that ``decides" to use only four blades, two of which are in correspondence to the 3- and 5-blade models. It is possible that, with further (or randomly different) training, the remaining two blades might coalesce to the third blade of those models. Again, this is weak evidence that the model's accuracy is intrinsic in the data, not in the model parameter details.

Since rows thus tend to be strongly assigned to one of three blades, one might naively guess that each blade specializes in one or more statistical subpopulations. Perhaps so, but it is worth remembering that the blades maximize their rows' one-minus predictability, not their rows' homogeneity. Table \ref{tab:PINP} below shows the responses for the question PERNP (personal earnings) stratified by blade. Since STUMS condenses the numerical answers to deciles, the percentage answers would all be 10\% if the blades were unspecialized.

\begin{table}[ht]
\centering
\caption{PERNP responses by blade (percents)}
\vspace{-6pt}
\label{tab:PINP}
\begin{tabular}{@{}lrrrrrrrrrr@{}} 
\toprule
& \multicolumn{10}{c}{decile in full sample} \\
\cmidrule(l){2-11}
blade & 1 & 2 & 3 & 4 & 5 & 6 & 7 & 8 & 9 & 10 \\ 
\midrule
1 & 7     & 15    & 9     & 10    & 8     & 13    & 7     & 14    & 7     & 11  \\
2 & 9     & 5     & 10    & 9     & 11    & 8     & 16    & 4     & 21    & 7   \\  
3 & 14    & 10    & 11    & 11    & 12    & 9     & 7     & 13    & 0     & 12  \\ 
\bottomrule
\end{tabular}
\end{table}

Especially in deciles 6-10, one sees the emergence of a kind of checkerboard pattern. Seemingly, by learning such a mask, the blade is better able to predict answers for those rows assigned to it. Each blade thus seems tuned to a large number of separate subpopulations whose predictions are non-interfering. One can imagine that the model is on the verge of discovering the concept of coded masks, as used in astronomy and elsewhere.\cite{Goldwurm} An equally unintuitive example is AGEP (age), shown in Table \ref{tab:AGEP}, where blade 3 inexplicably loves teenagers and hates twenty-somethings, while other age ranges are mostly uniform across the blades.

\begin{table}[ht]
\centering
\caption{AGEP responses by blade (percents)}
\vspace{-6pt}
\label{tab:AGEP}
\begin{tabular}{@{}lrrrrrrrrrrr@{}} 
\toprule
& \multicolumn{10}{c}{age range} \\
\cmidrule(l){2-12}
blade & n/a & 1-6 & 7-12 & 13-19 & 20-29 & 30-39 & 40-49 & 50-59 & 60-69 & 70-79 & 80+ \\ 
\midrule
1 & 0     & 8     & 7     & 9     & 19    & 11    & 10    & 13    & 13    & 7     & 3    \\ 
2 & 1     & 2     & 8     & 7     & 13    & 14    & 14    & 13    & 14    & 10    & 4    \\ 
3 & 2     & 8     & 8     & 14    & 2     & 14    & 13    & 12    & 13    & 9     & 5    \\
\bottomrule
\end{tabular}
\end{table}

\FloatBarrier

\section{Measuring Privacy}

\begin{figure}
\centering
\includegraphics[width=300pt]{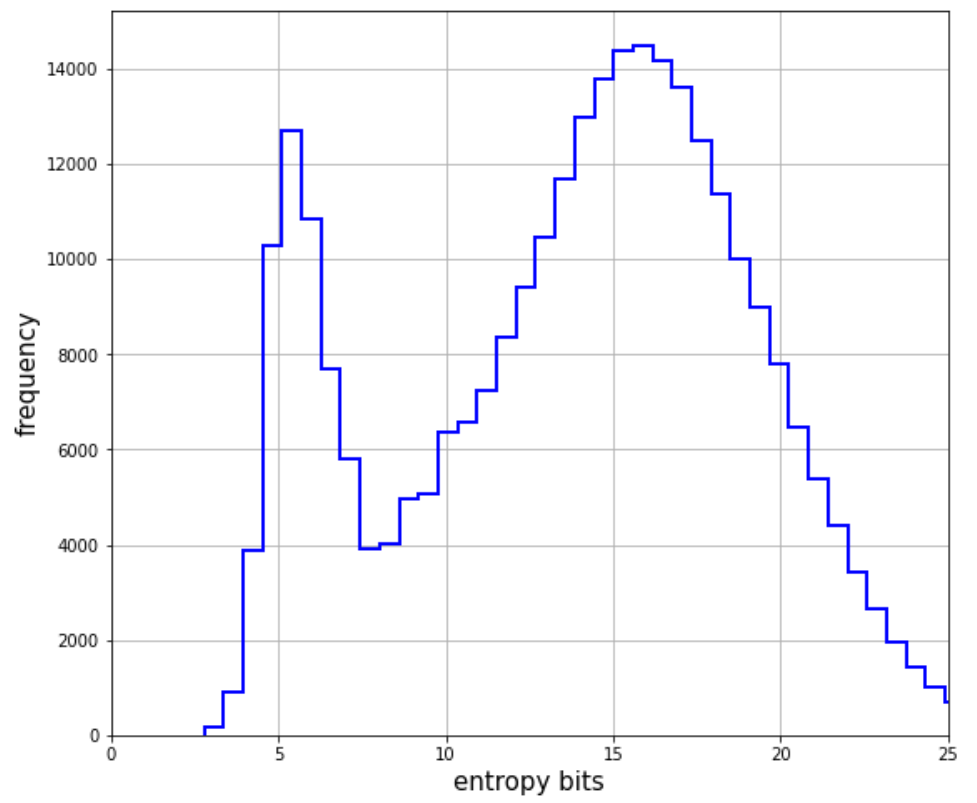}
\caption{\label{fig:entropybits}As each synthetic row is created, random draws are made from its predicted probability distribution for each question. Here plotted is the distribution over rows of the total entropy of each row's draws, as measured in bits. A bimodal distribution of children (left) vs.~adults (right) emerges.}
\end{figure}

We are not able to offer any rigorous mathematical guarantees, such as those provided by differential privacy \cite{Cummings2024Advancing}, for the MODP method as described in this paper. We can, however, offer some empirical measures of the privacy achieved. Our perspective is close to that of Bindschaedler et al.~\cite{bind2017}, who explore the concept of ``plausible deniability".

Each MODP synthetic row is created via the function \codefont{instantiate} (above, \S \ref{sec:NSPLR}) by independent random draws from probability distributions unique for each question and row. As coded, \codefont{instantiate} returns not just the synthetic row, but also the entropy of generating it.
As mentioned earlier, information theory tells us that, if $B$ bits of entropy are introduced, then a synthetic record is just one of about $2^B$ roughly equiprobable such records.\cite{shannon} 

Figure \ref{fig:entropybits} shows the distribution of entropies across all the rows in the sample. One sees an unexpected bimodal distribution that (after some inspection of the input data) turns out to be ``children" on the left and ``adults" on the right. That is, a given child is adequately specified by many fewer bits than an adult.

For adults, the typical synthetic record is one of $\sim 2^{16} \approx 6\times 10^4$ different possibilities. One of them might or might not be exactly a true record, but there is no causal reason for it to be so. This quantifies the plausible deniability.

What about children? Here, with just $\sim 6$ bits of specificity (see Figure \ref{fig:entropybits}), it is more likely that the synthetic record will exactly replicate a true one. But, by the same token, we might expect that such a true record occurs not once, but many times in the true dataset. This in itself provides plausible deniability.

\begin{figure}
\centering
\includegraphics[width=450pt]{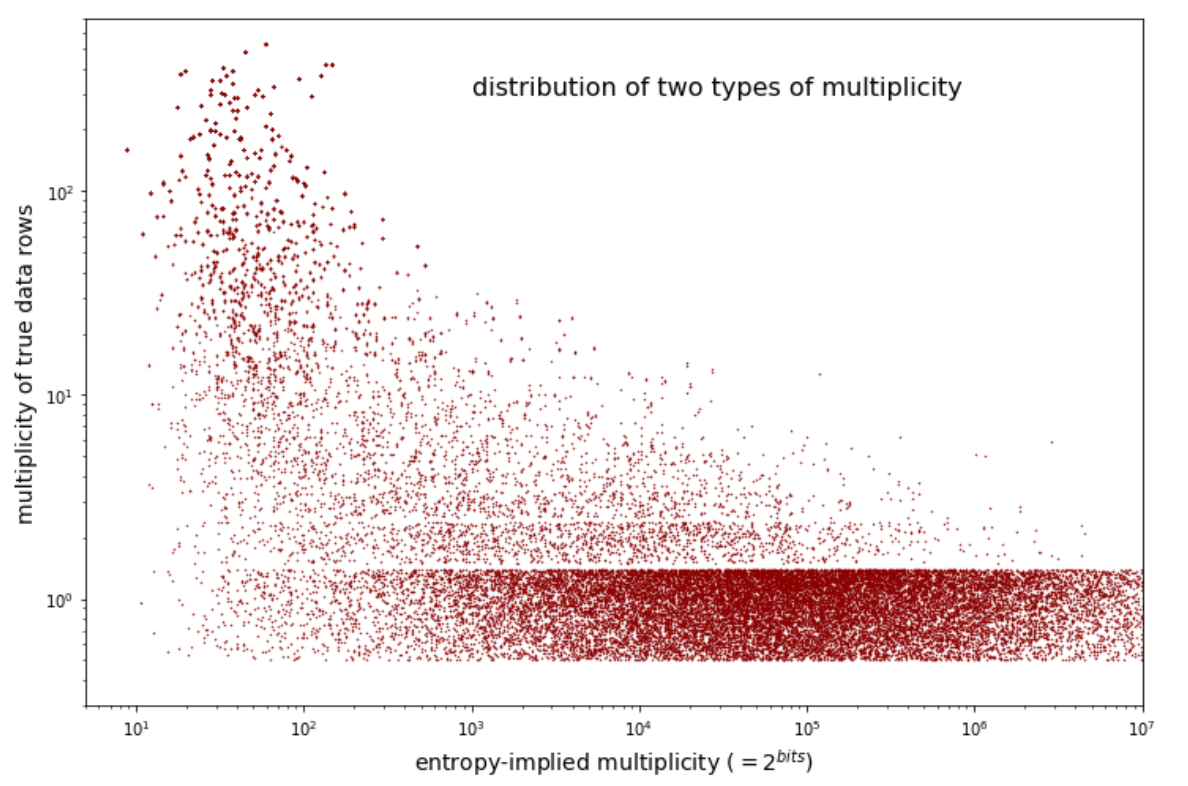}
\caption{\label{fig:twomultiplicity}Privacy can be protected either when a true row is identical to many other true rows (multiplicity) or when its corresponding synthetic row is created with a large entropy of random draws. The co-distribution is mainly children at the upper left, adults at the bottom and right. (For visibility, small integer $y$ values have been broadened $\pm 0.5$ into bands.}
\end{figure}

In other words, there are two kinds of multiplicity in which to hide: the multiplicity of the true record, and the multiplicity of generating a true record's synthetic partner. Figure  \ref{fig:twomultiplicity} shows a scatter plot of both kinds across all the rows. One sees that only a small number of rows are deficient in both kinds; if desired, these could be removed from the synthetic dataset with little impact.

\begin{figure}
\centering
\includegraphics[width=350pt]{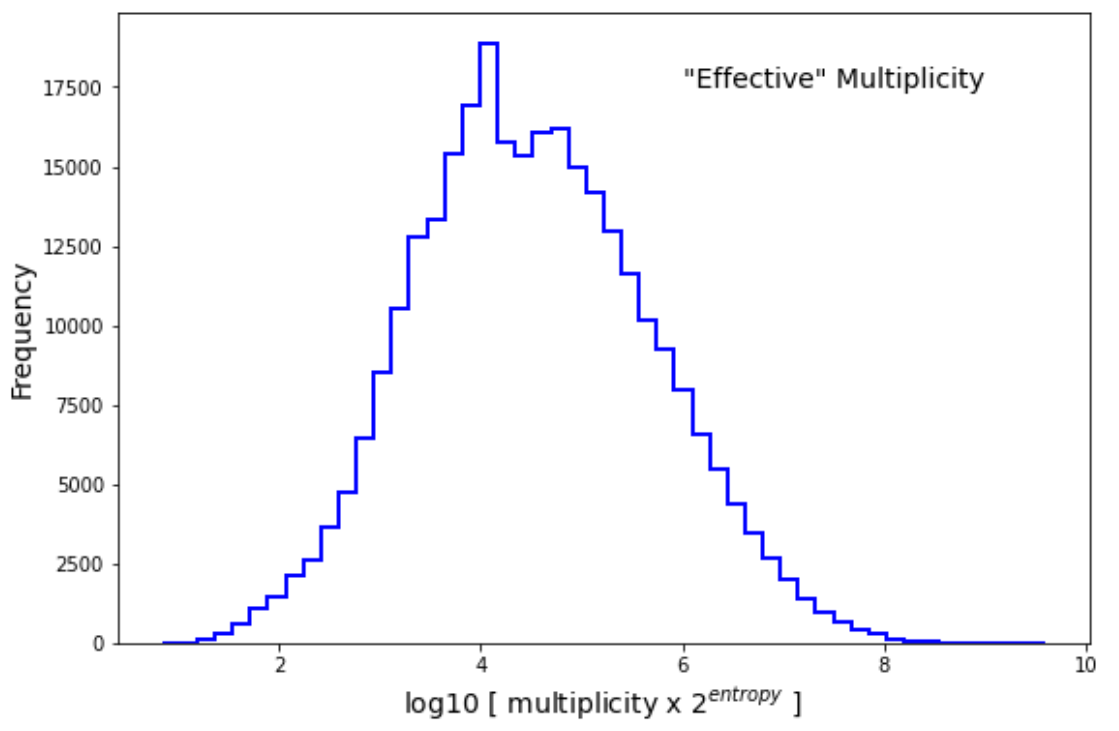}
\caption{\label{fig:effectivemultiplicity}An ``effective multiplicity" is the product of the two types of multiplicity shown in Figure \ref{fig:twomultiplicity}. It is seen to be generally large (median $\sim 10^4$) for both children and adults.}
\end{figure}

Purely heuristically, we can define an ``effective multiplicity" as the product of the two types of multiplicity, and take this as something like a measure of plausible deniability.
Figure \ref{fig:effectivemultiplicity} shows a histogram of this quantity. The distribution peaks around a value $10^4$. (Note, however, that the multiplicity of the input records decreases in proportion to their sample size.)

There remains one loophole that we need to eliminate by an explicit check. Although \codefont{instantiate} chooses one exemplar among $2^B$ possible (where $B$ is the entropy in bits), it is possible that the cloud of all the examplars does not have sufficient diameter to spread it over many input records. If that were so, then, while not identical, a true record might be identified by being the nearest neighbor (or among the few nearest neighbors) of a synthetic record.

\begin{figure}
\centering
\includegraphics[width=350pt]{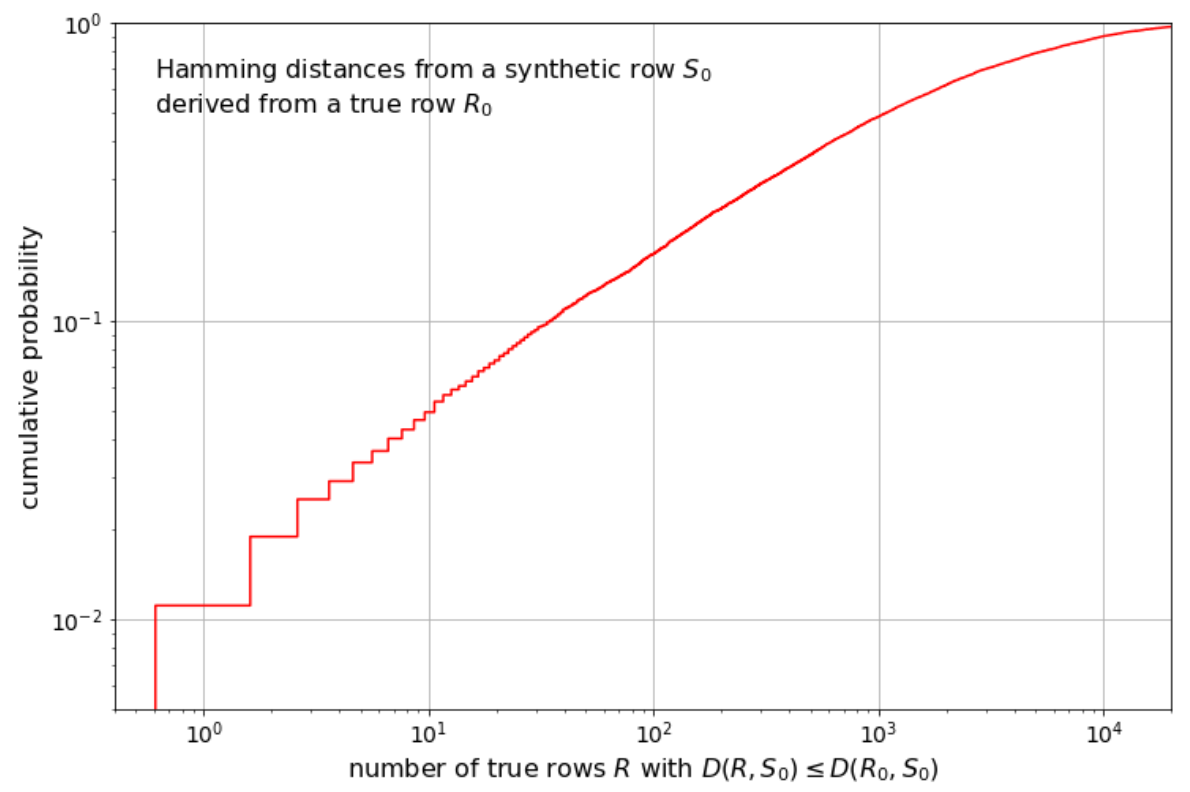}
\caption{\label{fig:hamming}Number of true rows as close or closer to a given synthetic rows than is its causal partner. The nearest true row to a given synthetic row is causal only about 1\% of the time in the STUMS dataset. The causal row is within the nearest 10 only about 5\% of the time and is typically lost among several thousand non-causals.}
\end{figure}

One is free to choose any metric for distance, but suppose that we choose Hamming distance (that is, the number of differing categorical columns). Figure \ref{fig:hamming} shows, for a large sample of rows, the number of true rows as close or closer to a given synthetic rows than is its causal partner---in other words, the number of rows contributing to plausible deniability. One sees that the nearest true row to a given synthetic row is causal only about 1\% of the time. The causal row is within the nearest 10 only about 5\% of the time and is typically lost among several thousand non-causals. Notably, the attacker cannot distinguish the causal 1\% or 5\% cases from the 99\% non-causal ones. Whether this degree of privacy by plausible deniability is good enough is a matter of policy, not statistics. 

\FloatBarrier
\section{Discussion}

STUMS, here investigated, comprises a subset of U.S. Census PUMS roughly a factor of 10 smaller in number of questions (data columns) and in number of samples (data rows), in comparison with the full U.S. sample. This subset requires only a very small neural-net MODP model, and needs only a desktop compute resource to train, at most hours on a single consumer-grade GPU.

Scaling to the full U.S.~PUMS should produce a model $\sim 100\times$ larger that can be evaluated, once trained, across all the PUMS data with $\sim 1000\times$ compute ops. This should be easily attainable on a production-level machine.

Less clear is how the compute requirements for training should scale. A guess is $10^4$ times, which might be realized as $\sim 100$ hours on a total of $\sim 100$ GPUs, parallelizing by the geographical subsets of the PUMS. Of course, one would not immediately scale up to this size without several intermediate experiments.

The accuracy of the approximation of conditional independence (equation \ref{eq:condind}) should get better, not worse, as the number of columns is increased, because new information is added, not subtracted (see Supplementary Materials). One might therefore hope for even better accuracy than that here demonstrated.

So as to make this paper an investigation on purely categorical data, STUMS converted numerical values such as income to categorical values, usually as deciles (see \S\ref{sec:STUMS}). Numerical questions generally allow more-, not less-, powerful simulation techniques, such as correlational models, Gaussian copulas, etc.; or numerical values could be more directly predicted by a neural net akin to, but more complicated than, MODP. Or, one could train initially on discretized quantiles, then sub-resolve each quantile with a separately trained continuous model.  Such variants are beyond the scope of this initial investigation.

\subsection*{Data availability}

All code utilized, and all the trained models mentioned, will be made available on GitHub at \codefont{https://github.com/whpress/SyntheticCategoricalData}.

\subsection*{Acknowledgments}

I thank Dan Meiron for discussions, and Jon Ullman for pointing me to relevant literature.

\bibliographystyle{unsrt}
\bibliography{sample}

\begin{thebibliography}{10}

\bibitem{ACS}
{United States Census Bureau}.
\newblock {American Community Survey (ACS)}.
\newblock {https://www.census.gov/programs-surveys/acs}, 2022.
\newblock Accessed: 2024-05-24.

\bibitem{PUMS}
{United States Census Bureau}.
\newblock {Public Use Microdata Sample (PUMS)}.
\newblock {https://www.census.gov/programs-surveys/acs/microdata.html}, 2022.
\newblock Accessed: 2024-05-24.

\bibitem{Ullman}
Jonathan Ullman and Salil Vadhan.
\newblock {PCPs} and the hardness of generating private synthetic data.
\newblock {\em Theory of Cryptography - 8th Theory of Cryptography Conference}, pages 400--416, 03 2011.

\bibitem{UN2022}
{United Nations Economic Commission for Europe}.
\newblock {\em Synthetic Data for Official Statistics: A Starter Guide}.
\newblock United Nations, Geneva, 2022.

\bibitem{Mathur2024}
Shirley Mathur, Yajuan Si, and Jerome~P. Reiter.
\newblock Fully synthetic data for complex surveys.
\newblock {\em arXiv}, 2309.09115v4, 4 2024.

\bibitem{jordon2018}
James Jordon, Jinsung Yoon, and Mihaela Van Der~Schaar.
\newblock {PATE-GAN}: Generating synthetic data with differential privacy guarantees.
\newblock In {\em Proceedings of the International Conference on Learning Representations}, pages 1--21, 2018.

\bibitem{zhang2014}
Jun Zhang, Graham Cormode, Cecilia~M. Procopiuc, Divesh Srivastava, and Xiaokui Xiao.
\newblock Privbayes: Private data release via {Bayesian} networks.
\newblock In {\em Proceedings of the 2014 {ACM SIGMOD} international conference on Management of data}, pages 1423--1434. ACM, 2014.

\bibitem{abay2018}
N.~C. Abay, Y.~Zhou, M.~Kantarcioglu, B.~Thuraisingham, and L.~Sweeney.
\newblock Privacy preserving synthetic data release using deep learning.
\newblock In {\em Proceedings of the Joint European Conference on Machine Learning and Knowledge Discovery in Databases}, pages 510--526. Springer, 2018.

\bibitem{Efron81}
{Wikipedia}.
\newblock Bootstrapping (statistics).
\newblock [Online; accessed 7-June-2024].

\bibitem{shannon}
Claude~E. Shannon and Warren Weaver.
\newblock {\em The Mathematical Theory of Communication}.
\newblock University of Illinois Press, Urbana, IL, 1949.

\bibitem{PyTorch}
Adam Paszke and et~al.
\newblock Pytorch: An imperative style, high-performance deep learning library.
\newblock In {\em Advances in Neural Information Processing Systems 32}, pages 8024--8035. Curran Associates, Inc., 2019.

\bibitem{attention}
Ashish Vaswani, Noam Shazeer, Niki Parmar, Jakob Uszkoreit, Llion Jones, Aidan~N. Gomez, Lukasz Kaiser, and Illia Polosukhin.
\newblock Attention is all you need.
\newblock {\em CoRR}, abs/1706.03762, 2017.

\bibitem{NIST}
{National Institute of Standards and Technology (NIST)}.
\newblock {Engineering Statistics Handbook - Process Capability, \S 7.3.3}.
\newblock https://www.itl.nist.gov/ div898/handbook/prc/section3/prc33.htm.
\newblock Accessed: 2024-05-24.

\bibitem{Warner1965}
S.~L. Warner.
\newblock Randomised response: a survey technique for eliminating evasive answer bias.
\newblock {\em Journal of the American Statistical Association}, 60(309):63--69, March 1965.

\bibitem{Greenberg1969}
B.~G. Greenberg et~al.
\newblock The unrelated question randomised response model: Theoretical framework.
\newblock {\em Journal of the American Statistical Association}, 64(326):520--539, June 1969.

\bibitem{pseudocount}
{Wikipedia}.
\newblock Additive smoothing: Pseudocount.
\newblock [Online; accessed 7-June-2024].

\bibitem{Goldwurm}
Andrea Goldwurm and Aleksandra Gros.
\newblock Coded mask instruments for gamma-ray astronomy.
\newblock {\em arXiv}, 2305.10130v1, 5 2023.

\bibitem{Cummings2024Advancing}
Rachel Cummings, Damien Desfontaines, David Evans, Roxana Geambasu, Yangsibo Huang, Matthew Jagielski, and {et al.}
\newblock {Advancing} {Differential} {Privacy}: Where {We} {Are} {Now} and {Future} {Directions} for {Real}-{World} {Deployment}.
\newblock {\em Harvard Data Science Review}, 6(1), Jan 16 2024.
\newblock https://hdsr.mitpress.mit.edu/pub/sl9we8gh.

\bibitem{bind2017}
Vincent Bindschaedler, Reza Shokri, and Carl~A. Gunter.
\newblock Plausible deniability for privacy-preserving data synthesis.
\newblock {\em Proceedings of the VLDB Endowment}, 10(5):481--492, 2017.

\end{thebibliography}

\section*{Supplementary Materials}

The American Community Survey Public Use Microdata Sample (PUMS) makes available samples of two different types of records. Our STUMS simplified subset (\S\ref{sec:STUMS}) was based on PUMS ``person records" and their associated questions. As a separate check on the generality of this paper's MODP data synthesis method, we can generate a separate simplified subset based on PUMS ``housing records" and its different set of questions. We term this subset STUMS-H.

As for STUMS, STUMS-H comprises geographically all Texas responses. STUMS-H has 133,016 rows and 46 questions that expand to 315 binary columns.
The following table, analogous to Table \ref{Table1}, lists the questions included.

\begin{longtable}{@{} l >{\raggedright\arraybackslash}p{6cm} ccc @{}}
\caption{STUMS-H Subset of PUMS Questions} \label{stums-h} \\
\toprule
Question & Description & \# Cats & Merged & Quantiled \\
\midrule
\endfirsthead
\multicolumn{5}{c}{{Table \thetable\ (continued)}} \\
\toprule
Question & Description & \# Cats & Merged & Quantiled \\
\midrule
\endhead
\midrule
\endfoot
\bottomrule
\endlastfoot
NP & Number of persons in this household & 8 & Yes & \\ 
NPF & Number of persons in family (unweighted) & 8 & Yes & \\ 
ACR & Lot size & 4 & & \\ 
NR & Presence of nonrelative in household & 3 & & \\ 
YRBLT & When structure first built & 6 & Yes & \\ 
TEN & Tenure & 5 & & \\ 
VALP & Property value & 11 & & Yes \\ 
RMSP & Number of rooms & 7 & Yes & \\ 
HFL & House heating fuel & 10 & & \\ 
WATP & Water cost  & 11 & & Yes \\ 
BLD & Units in structure & 11 & & \\ 
RNTP & Monthly rent  & 11 & & Yes \\ 
GRNTP & Gross rent  & 11 & & Yes \\ 
MHP & Mobile home costs (yearly amount) & 12 & & Yes \\ 
MRGP & First mortgage payment (monthly amount) & 11 & & Yes \\ 
SMP & Total payment on second mortgages and home equity loans (monthly) & 12 & & Yes \\ 
SMOCP & Selected monthly owner costs & 12 & & Yes \\ 
INSP & Fire/hazard/flood insurance (yearly amount) & 12 & & Yes \\ 
TAXAMT & Property taxes (yearly real estate taxes) & 12 & & Yes \\ 
PLM & Complete plumbing facilities & 3 & & \\ 
BATH & Bathtub or shower & 3 & & \\ 
BDSP & Number of bedrooms & 7 & Yes & \\ 
KIT & Complete kitchen facilities & 3 & & \\ 
VEH & Vehicles (1 ton or less) available & 8 & & \\ 
TEL & Telephone service & 3 & & \\ 
ACCESSINET & Access to the Internet & 4 & & \\ 
SMARTPHONE & Smartphone & 3 & & \\ 
LAPTOP & Laptop or desktop & 3 & & \\ 
BROADBND & Cellular data plan & 3 & & \\ 
COMPOTHX & Other computer equipment & 3 & & \\ 
HISPEED & Broadband Internet service  & 3 & & \\ 
HINCP & Household income (past 12 months) & 10 & & Yes\\ 
FINCP & Family income (past 12 months) & 10 & & Yes\\ 
FS & Yearly food stamp/SNAP recipiency & 3 & & \\ 
ELEP & Electricity cost (monthly cost) & 8 & Yes & \\ 
HHT & Household/family type & 8 & & \\ 
WIF & Workers in family during the past 12 months & 5 & & \\ 
PARTNER & Unmarried partner household & 6 & & \\ 
HUPAC & HH presence and age of children & 5 & & \\ 
HUGCL & Household with grandparent living with grandchildren & 3 & & \\ 
CPLT & Couple Type & 5 & & \\ 
HHLDRAGEP & Age of the householder & 9 & Yes & \\ 
HHLDRRAC1P & Recoded detailed race code of the householder & 6 & & \\ 
HHL & Household language & 6 & Yes & \\ 
LNGI & Limited English speaking household & 3 & & \\ 
FPARC & Family presence and age of related children & 5 & & \\ 
\midrule
Questions: 46 &   \multicolumn{2}{r}{Categories: 315
$\;$}&  & \\
\end{longtable}

\begin{figure}
\centering
\hspace*{-40pt}
\includegraphics[width=270pt]{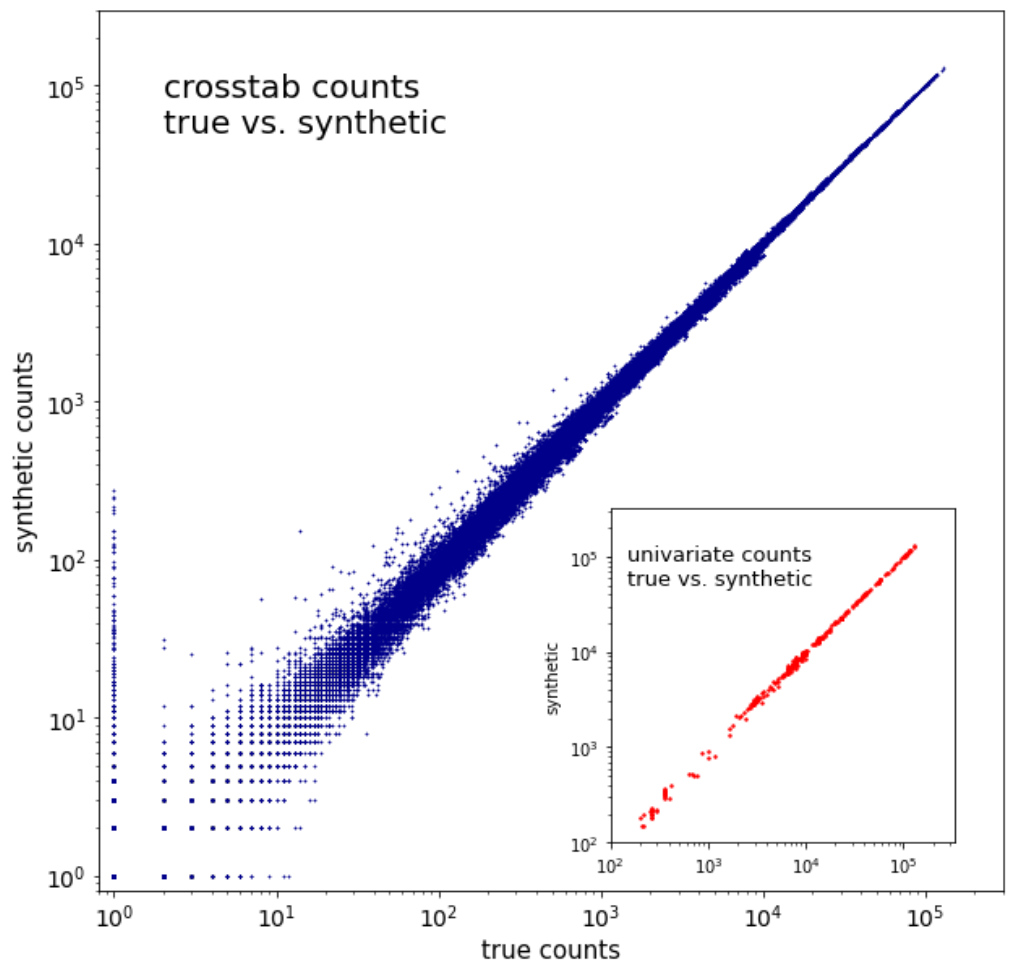}\includegraphics[width=270pt]{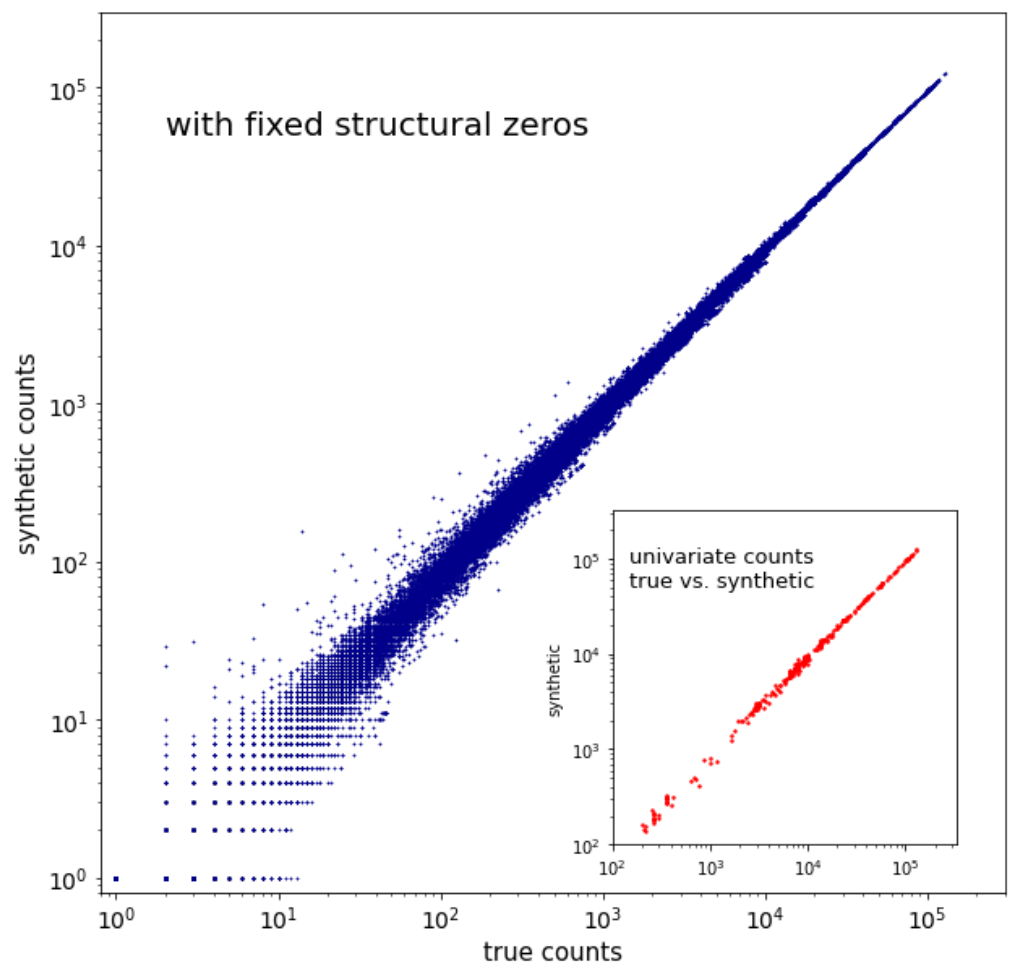}
\caption{\label{fig:h-fiveblade}Results from a 5-blade model trained on the STUMS-H dataset. STUMS-H has 133,016 rows (households, not persons) and 46 questions (different from STUMS questions) that expand to 315 binary columns. The MODP method is similarly successful with both STUMS and STUMS-H.}
\end{figure}

We define a model with 5 blades and 15 reduced features. As with STUMS, we first train with a loss function \codefont{torch.nn.MSELoss}, then switch to \codefont{zval\_loss\_function}, above.

Figure \ref{fig:h-fiveblade} shows results for STUMS-H analogous to Figures \ref{fig:fiveblade} and \ref{fig:tweaks} for STUMS. 

STUMS-H has substantially fewer rows than STUMS, and substantially more questions (or binary columns). It is not obvious a priori whether this should produce a more- or a less-accurate synthetic dataset.
The fewer rows may enable only less-accurate learning of the function ${\cal L}$, but the more questions may make the conditional independences (equation \ref{eq:condind}) more accurate. In the event,
the two datasets end up quite comparable. With the post-processing described in (\S \ref{sec:postp}), STUMS-H achieves a median fractional error of
4.1\% across $315\times (315+1)/2$ crosstab cells, as compared to STUMS' 
3.7\% across $233\times (233+1)/2$ cells. Table \ref{tab:HLogDevAcc}, the analog of Table \ref{tab:LogDevAcc}, shows several measures of crosstab fractional accuracy for the model and some of the same variants tried for STUMS.

\begin{table}[h]
\centering
\caption{Fractional (i.e., Log) Accuracy on STUMS-H Data}
\vspace{-6pt}
\label{tab:HLogDevAcc}
\begin{tabular}{@{}lcccl@{}} 
\toprule
& \multicolumn{4}{c}{accuracy over $315\times (315+1)/2$ crosstab cells} \\
\cmidrule(l){2-5}
model & median & mean absolute & r.m.s. & comment\\ 
\midrule
h-model\_5\_15 & 0.046 & 0.140 & 0.316 & as trained \\
h-model\_5\_15 & 0.070 & 0.151 & 0.249 & struct. zeros fixed \\
h-model\_5\_15 & 0.041 & 0.131 & 0.302 & post-processed (\S \ref{sec:postp})\\
h-model\_5\_15 (R.R.) & 0.030 & 0.104 & 0.252 & \S\ref{sec:RR} ($p=1/3$)\\
h-model\_5\_15 (R.R.) & 0.022 & 0.086 & 0.218  & \S\ref{sec:RR} ($p=1/2$)\\
ideal goal & 0.014 & 0.054 & 0.130 & resample of STUMS-H data\\
\bottomrule
\end{tabular}
\end{table}

\end{document}